\documentclass[aps,prd,10pt,notitlepage,nofootinbib,superscriptaddress,showkeys,showpacs]{revtex4-1}
\usepackage{amsfonts, amsmath, amssymb, amsthm}
\usepackage{color}
\usepackage{braket}
\usepackage{graphicx}

\newcommand{\be}{\begin{equation}}
\newcommand{\ee}{\end{equation}}
\newcommand{\bea}{\begin{eqnarray}}
\newcommand{\eea}{\end{eqnarray}}

\newcommand{\cA}{\mathcal{A}}
\newcommand{\cM}{\mathcal{M}}
\newcommand{\cB}{\mathcal{B}}
\newcommand{\cF}{\mathcal{F}}
\newcommand{\cC}{\mathcal{C}}
\newcommand{\cE}{\mathcal{E}}

\newcommand{\cG}{\mathcal{G}}
\newcommand{\cL}{\mathcal{L}}

\newcommand{\cT}{\mathcal{T}}

\newcommand{\prf}{{\noindent \bf Proof\; \; }}

\DeclareMathOperator{\tr}{Tr}

\newcommand{\cD}{{\cal D}}

\newtheorem{lemma}{Lemma}
\newtheorem{definition}{Definition}
\newtheorem{theorem}{Theorem}

\newtheorem{prop}{Proposition}

\begin{document}

\title{\Large Enhancing non-melonic triangulations: A tensor model mixing melonic and planar maps}

\author{{\bf Valentin Bonzom}}\email{bonzom@lipn.univ-paris13.fr}
\affiliation{LIPN, UMR CNRS 7030, Institut Galil\'ee, Universit\'e Paris 13, Sorbonne Paris Cit\'e, 99, avenue Jean-Baptiste Cl\'ement, 93430 Villetaneuse, France, EU}

\author{{\bf T.~Delepouve}}\email{delepouve@cpht.polytechnique.fr}
\affiliation{Laboratoire de Physique Th\'eorique, CNRS UMR 8627, Universit\'e Paris XI, 91405 Orsay Cedex, France, EU}
\affiliation{Centre de Physique Th\'eorique, CNRS UMR 7644, \'Ecole Polytechnique, 91128 Palaiseau Cedex, France, EU}

\author{{\bf V.~Rivasseau}}\email{rivass@th.u-psud.fr}
\affiliation{Laboratoire de Physique Th\'eorique, CNRS UMR 8627, Universit\'e Paris XI, 91405 Orsay Cedex, France, EU}
\affiliation{Perimeter Institute for Theoretical Physics, 31 Caroline St. N, ON N2L 2Y5, Waterloo, Canada}

\begin{abstract} Ordinary tensor models of rank $D\geq 3$ are dominated at large $N$ by tree-like graphs, known as melonic triangulations. We here show that non-melonic contributions can be enhanced consistently, leading to different types of large $N$ limits. We first study the most generic quartic model at $D=4$, with maximally enhanced non-melonic interactions. The existence of the $1/N$ expansion is proved and we further characterize the dominant triangulations. This combinatorial analysis is then used to define a non-quartic, non-melonic class of models for which the large $N$ free energy and the relevant expectations can be calculated explicitly. They are matched with random matrix models which contain multi-trace invariants in their potentials: they possess a branched polymer phase and a 2D quantum gravity phase, and a transition between them whose entropy exponent is positive. Finally, a non-perturbative analysis of the generic quartic model is performed, which proves analyticity in the coupling constants in cardioid domains.
\end{abstract}

\noindent  Pacs numbers: 02.10.Ox, 04.60.Gw, 05.40-a
\keywords{Random tensors, Regular edge-colored graphs, planar maps}

\maketitle

\section{Introduction}

An interesting way to define and work with continuous random geometries (for instance, the continuous random tree) is to start with random discrete spaces (for instance, trees) whose sizes can become arbitrarily large. In a suitable scaling limit, such random discrete objects can converge in a precise (Gromov-Hausdorff-Prokhorov \cite{Gromov,AbDelHos}, or GHP for short) sense to a continuous random space of a certain universality class.

It is standard for physicists to think of discrete random spaces as combinatorial objects generated by the Feynman expansion of some integrals over tensors. A tensor of rank $D$ is a $D$-dimensional array of (complex) numbers. A position in such an array is labeled by a $D$-uple of integers each ranging from $1$ to $N$. Of particular interest to us are random tensors equipped with a $U(N)^D$-invariant distribution. If the Feynman expansion can be organized as a series in $1/N$, then the large $N$ limit exists and selects a class of combinatorial objects (the random spaces) to be summed over.

A tensor of rank 1 is a vector and the large $N$ limit of vector models consists of plane trees (in the intermediate field representation) whose GHP limit is the continuous random tree (CRT, for short) \cite{aldous}. The CRT, also known as the branched polymer universality class, has Hausdorff dimension 2 and spectral dimension 4/3.

The situation changes drastically for tensors of rank two, i.e. random matrices. Indeed, matrix models provide a quantization of two-dimensional gravity coupled to conformal matter \cite{matrix,Kazakov,mm}. Such matrix integrals typically generate maps (possibly decorated with matter) weighted by $N^{2-2g}$ where $g$ is the genus of the map \cite{Hooft}. This topological expansion is dominated at large $N$ by planar maps (of spherical topology). As random spaces, planar maps (without critical matter) converge in the GHP sense to the Brownian sphere \cite{LeGall,LeGallMiermont}, a random space with Hausdorff dimension 4 and (expected) spectral dimension 2.

In higher dimensions matrix models generalize to tensor models \cite{oldtens1,oldtens2} and group field theories \cite{boulatov,laurentgft,Krajewski:2012aw,Oriti:2013aqa}. However such models did not admit a large $N$ limit. The situation changed with the advent of colored \cite{color,review} and $U(N)^{D}$-invariant \cite{uncoloring} tensor models. They admit a $1/N$ expansion \cite{expansion1,expansion2,expansion3}.  
This perturbative Feynman expansion furthermore sums over all manifolds of dimension $D$ and a restricted class of pseudo-manifolds \cite{lost}. The weights of those objects are moreover simple discretizations of the Einstein-Hilbert action \cite{ambjorn}, coinciding with Euclidean dynamical triangulations. Hence tensor models seem relevant for the quantization of gravity in dimensions higher than 2 \cite{Rivasseau:2011hm,Rivasseau:2012yp,Rivasseau:2013uca}. They can support matter fields \cite{IsingD,b01,b02,b03} and have statistical mechanics applications \cite{spinglass}. 

The first analytic solution of tensor models at large $N$, for $D\geq 3$, matched the expectations from the numerical analysis of Euclidean dynamical triangulations, \cite{critical, uncoloring}. Mainly, they are dominated by melonic graphs which are representatives of the universality class of branched polymers \cite{BP}. The coupling to matter reproduces the critical exponents of multi-critical branched polymers. Notice that the colored and invariant families of tensor models admit double-scaling limits for $D<6$ \cite{KOR,Dartois:2013sra,GS,DSSD}, which are also made of trees. We would like to point out that trees can serve as building blocks for more elaborated objects. A key idea is to decorate them cleverly with a certain density of loops, for instance by additional data such as Schaeffer's ``well-labelings'' \cite{schaeffer}  (the process which leads to the Brownian sphere), or by mating two trees (a process which leads to a kind of Brownian sphere decorated with an SLE called a ``peanosphere'' \cite{Duplantier:2014daa}). The present article precisely shows a way for tensor models to go from trees to planar maps through a phase transition which exhibits a proliferation of planar maps along a tree.

There is indeed more to tensor models than melonic graphs. For instance, one can trivially build a square matrix from a tensor of rank 4 and equip it with an ordinary matrix model, which would converge at large $N$ to the Brownian sphere. Importantly, this requires the distribution on the random variables to have a different behavior with $N$ than those dominated by melonic graphs. Furthermore, the boundary observables of tensor models are far more richer than those of matrix models: instead of collections of loops, they are labeled by $D$-regular graphs. Most of them usually contribute to sub-leading corrections, even when they explicitly enter the measure. However, it might be possible to modify the behavior of the measure with $N$ so as to enhance such non-melonic observables. This idea was first developed in \cite{b2}. It showed\footnote{This analysis was also required in order to define tensor models for ``rectangular'' tensors.} that while some models with non-melonic distributions simply extend the range of applicability of the universality theorem of \cite{universality}, some other models can escape it and converge to some random space that is not the CRT.

It was in fact noticed in the last section of \cite{universality} that every tensor observable which appears in the measure should admit a \emph{range of admissible scalings} with $N$ for which the $1/N$ expansion exists. For a melonic invariant, this range is restricted to a single possible value, which is the one used in \cite{uncoloring, universality}. But non-melonic invariants can lead to different types of $1/N$ expansion, as established and exploited in \cite{b2}. The present article pushes this approach further: we start the analysis of a new kind of $1/N$ expansion for tensor models. It is based on melonic interactions, and non-melonic ones which in contrast to the existing literature (except for \cite{b2}) are appropriately rescaled to their largest admissible scaling. We say that those observables are enhanced and call such models \emph{maximally rescaled}.

In Section \ref{sec:max}, we explain the notion of maximally rescaled observable and briefly review basic facts about tensor models (including the structure of the Feynman graphs and a comparison of the present approach with \cite{b2}).

In Section \ref{sec:quartic}, the most generic quartic, maximally rescaled model at $D=4$ is analyzed. The combinatorial analysis is performed thanks to the intermediate field method. It proves the existence of the $1/N$ expansion and characterizes the graphs which dominate the free energy at large $N$.

In Section \ref{sec:NecklaceTrees}, we introduce another generic class of maximally rescaled models whose measures involve non-melonic, non-quartic invariants which we call \emph{trees of necklaces} (still at $D=4$). The motivation is that these observables are boundary observables of the dominant graphs of one of the quartic models. This makes possible to use the combinatorial analysis of Section \ref{sec:quartic} to identify the graphs which dominate at large $N$. Combined with Schwinger-Dyson equations (or Tutte equations), that leads to their enumeration which allows the large $N$ evaluation of the free energy (in particular its critical entropy exponents) and expectations of observables. While the critical behaviors which arise in this class of models are known from the literature on matrix models with multi-trace invariants \cite{AlvarezBarbon, Das, KlebanovHashimoto, Korchemsky, BarbonDemeterfi}, our analysis paves the way for the future analysis of more complicated maximally rescaled models which would better take advantage of the richness of tensor observables.

Finally, Section \ref{sec:analytic} offers a non-perturbative analysis of the quartic maximally-rescaled models for any rank $D$. It establishes uniform analyticity and Borel summability at large $N$ in cardioid domains of the couplings. That is a non-trivial extension of the results of \cite{Delepouve:2014bma} which establishes similar results for quartic models of any rank, but without enhancement of non-melonic observables.

\section{Models with maximally enhanced interactions}\label{sec:max}

\subsection{Maximally enhanced observables}

Consider a Hermitian inner product space $V$ of dimension $N$ and $\{e_n| n=1,\dotsc, N\}$ an orthonormal basis in $V$. 
The inner product allows us to identify the dual of $V$ and its complex conjugate $\bar V$, hence not to distinguish between $\{e_n| n=1,\dotsc, N\}$ and its dual basis, nor between upper and lower indices. A complex tensor of rank $D$ is a multilinear form ${\bf T}:  V^{\otimes D} \to \mathbb{C}$, which can be written in components as
\be
{\bf T} = \sum_{n_1,\dotsc, n_D} T_{n_1\dotsb n_D} \; \;e_{n_1} \otimes \dotsb \otimes e_{n_D} \; .
\ee
It is important that $T_{n_1\dotsb n_D}$ has no symmetry properties. Hence its indices have a well defined position. 
We call the position of an index its \emph{color}. We have therefore a set $\cD$ of $D$ colors. From now on
we specialize to rank $D=4$ hence to $\cD =\{1,\dots 4\}$, and ${\bf T} = \sum_{n_1,\dots n_4} T_{n_1\dotsb n_4} \; \;e_{n_1} \otimes \dotsb \otimes e_{n_4}$. The dual tensor $  \overline{{\bf T} }$ is defined by 
\be 
\overline{{\bf T}} 
= \sum_{n^1, \dotsc, n^4} \overline{T}_{n_1\dotsb n_4}\; \; e_{n_1} \otimes \dotsb\otimes e_{n_4} \; .
\ee

The unique quadratic invariant is the (scalar) Hermitian pairing of $ \overline{{\bf T}}$ and ${\bf T}$ which 
writes:
\be
 \overline{{\bf T}}\cdot_{\cD  }  {\bf T} = \sum_{n_1, n_2, n_3, n_4}   \overline{T}_{n_1\dotsb n_4} {T}_{n_1\dotsb n_4}  \; . \label{uniquequadra}
\ee

Observables are monic, homogeneous polynomials invariant under $U(N)^D$. Given a finite set of such observables, $\{B_i({\bf T},\overline{{\bf T}})\}_{i\in I}$, we define the action $S$, the partition function $Z$ and the free energy $F$ as 
\begin{equation} \label{FreeEnergy}
\begin{aligned}
S({\bf T}, \overline{{\bf T}}) &= \overline{{\bf T}}\cdot_{\cD} {\bf T} + \sum_{i\in I} t_i B_i({\bf T}, \overline{{\bf T}}),\\
\exp (-F) &= Z = \int \prod_n \frac{d\overline{T}_n dT_n}{2\imath\pi}\ \exp \Bigl(-N^\alpha S({\bf T}, \overline{{\bf T}})\Bigr) .
\end{aligned}
\end{equation}
The exponent $\alpha$ which parametrizes the measure is usually fixed by the following requirements:
\begin{itemize}
\item the free energy is bounded, for any choice of $\{B_i\}_{i\in I}$, by a polynomial bound at large $N$ of type $|F| \leq N^{K}$,
\item there exist interactions $\{B_i\}_{i\in I}$ so that expanding the free energy perturbatively in the coupling constants around the Gaussian measure, there are infinitely many Feynman graphs with the same exponent of $N$,
\item the coupling constants $\{t_i\}_{i\in I}$ are independent of $N$.
\end{itemize}
These criteria uniquely select $\alpha = D-1$. The first two requirements are crucial. The first one ensures that the free energy, appropriately rescaled like $F/N^K$, is not divergent at large $N$. This requires $\alpha$ not to be too small. The second one is necessary to avoid trivial models where only a finite number of Feynman graphs would contribute at a fixed order of the $1/N$ expansion. This requires $\alpha$ not to be too large.

However, the last requirement does not seem equally important. In fact, it can be relaxed to the benefit of getting new large $N$ behaviors, as we will show in the present article. The situation is slightly similar to the case of matrix models. There $D=2$ and $\alpha=1$ for actions which are sums of single-trace invariants. Multi-trace invariants can be included provided they come with $N$-dependent weights, otherwise the free energy is divergent.

To see how that works, we need to give a bit more information on the usual large $N$ limit of tensor models. The free energy goes like $N^D$ and it is dominated by the melonic observables, i.e. $F$ only depends at large $N$ on the coupling constants of melonic polynomials. In other words, the second criterion above is only satisfied for melonic interactions.

If $B$ is non-melonic, the $B$-action $S_B = \overline{{\bf T}}\cdot_{\cD} {\bf T} + t_B\,B({\bf T}, \overline{{\bf T}})$ leads to a free energy $F_B$ which is independent of $t_B$ at large $N$. However, there may exist an $N$-dependent weight $t_B = N^{\alpha_B} \tilde{t}_B$, with $\tilde{t}_B$ $N$-independent, such that the associated $B$-action preserves the boundedness of the free energy while leading to a non-trivial (and non-melonic) model. If such an $\alpha_B >0$ exists, we say that $B$ can be maximally enhanced. Here ``maximally'' refers to the fact that if the value of $\alpha_B$ is increased, boundedness of the free energy is lost.

It is quite simple to see that such models do exist. For instance, set $D=4$ and consider ${\bf T}$ as a matrix of size $N^2\times N^2$ between pairs of tensor indices. Then, matrix model building applies as usual (and leads to a 2D quantum gravity phase at large $N$).

This was already observed in \cite{b2}, with a different but equivalent strategy. Instead of keeping $\alpha=D-1$ fixed and enhancing non-melonic polynomials, it was chosen to decrease $\alpha<D-1$ and simultaneously add some $N$-dependent weights on melonic observables so that $F$ does not diverge. It is easy to check that upon a rescaling of ${\bf T}$, this is equivalent to the approach we have just described.

However, our present work improves \cite{b2} on many aspects. For instance, the models described in \cite{b2} are labeled by ``slices of colors'', i.e. a choice of partition of the $D$ indices of the random tensor. As a consequence, observables which only differ by a permutation of some tensor indices would receive in \cite{b2} different $N$-dependent weights. There is no such partition here and all observables which differ by a relabeling of the indices receive the same enhancing.

Furthermore, in the context of the color slices, it was proved in \cite{b2} that the large $N$ limit is always of the same type as in tensor models with their standard scaling, except if the action contains some generalized matrix-like observables\footnote{These observables are defined technically as having a single face of colors $(i,j)$ and a single face of colors $(k,l)$, where $\{i,j,k,l\} = \{1,2,3,4\}$.}. However, the large $N$ limit of the latter models was not studied, and neither were models containing both melonic and matrix-like interactions. In the present article, we will give a much more comprehensive analysis of what happens when melonic observables are mixed with certain matrix-like observables. This mixing means that those observables will all appear in Feynman graphs. We will then see that this can be used to include new observables which result from gluing melonic and matrix-like observables together in the action of the model.

\subsection{Feynman Graphs}

\begin{figure} 
\includegraphics[scale=.5]{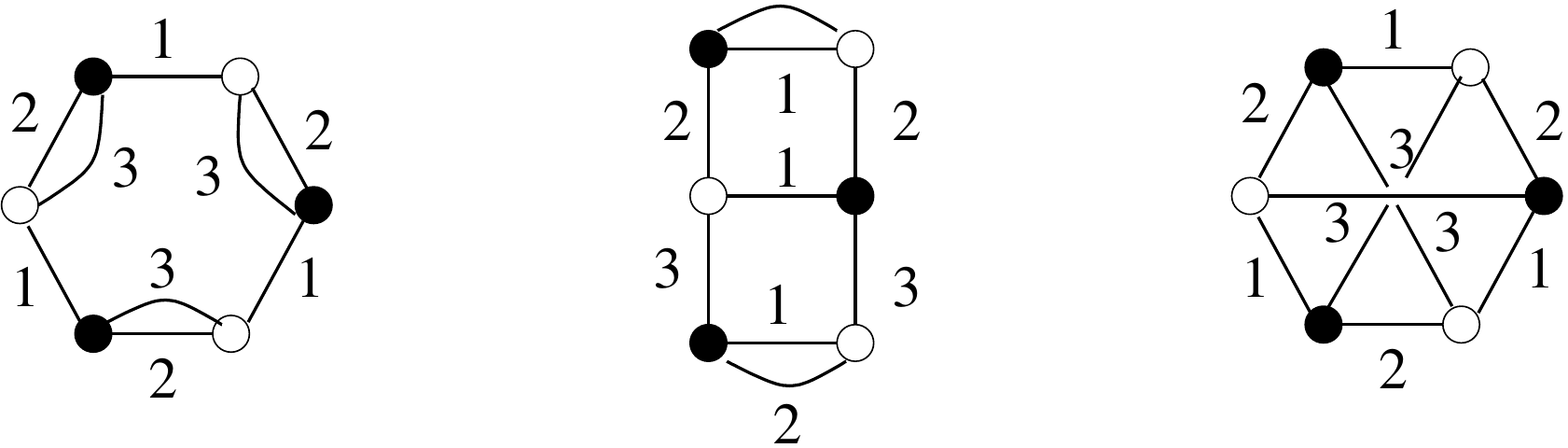}
\caption{The seven invariants at rank 3 and order 6: there are two different melonic families each with three members (shown left and center) and 
the non-planar bipartite complete graph $K_{3,3}$ (shown right).}\label{order6}
\end{figure}

The ordinary perturbative expansion is obtained by expanding the exponential of the non-quadratic terms of $S$ in \eqref{FreeEnergy} (called the interactions of the theory) into a series in the couplings $\{t_i\}_{i\in I}$ then (illegally!\footnote{The legal version of this procedure is constructive theory, see Section \ref{sec:analytic}.}) commuting the sum of the series with the remaining Gaussian integral. The resulting series is indexed by Wick contractions, also called \emph{Feynman graphs} which we now describe.

First, we recall that invariant polynomials can be represented as bipartite, $D$-regular graphs with edges colored in $\{1,\dotsc, D\}$ such that no two edges incident to the same vertex carry the same color. A white (or black) vertex represents ${\bf T}$ (or $\overline{{\bf T}}$) and an edge of color $c$ between two of them represents the contraction of their indices in position $c$. We call such graphs \emph{bubbles}. Some bubbles are on display in Figure \ref{order6}.

The Feynman graphs generated by tensor integrals consist of bubbles chosen in arbitrary (finite) numbers among those which represent the invariant polynomials $\{B_i\}_{i\in I}$, connected via additional edges to which the color 0 is attributed. The resulting graphs are bipartite, $(D+1)$-regular graphs whose edges have a color in $\{0,\dotsc,D\}$ such that the edges incident to a given vertex all have different colors. We call them vacuum Feynman graphs and the free energy expands as a sum over the connected ones.

If $\cG$ is a connected vacuum Feynman graph generated by the logarithm of the integral \eqref{FreeEnergy}, containing $n_i(\cG)$ bubbles of type $B_i$, for each $i\in I$, then the amplitude associated to $\cG$ reads
\begin{equation} \label{Amplitude}
A(\cG) = N^{F(\cG) - \alpha E_0(\cG) + \alpha \sum_{i\in I} n_i(\cG)}\, \prod_{i\in I} (-t_i)^{n_i(\cG)},
\end{equation}
which also works in the maximally rescaled case where the couplings $t_i$ may have an $N$-dependence. Here $E_0(\cG)$ is the number of edges of color 0 (it equals the number of white vertices of $\cG$). The quantity $F(\cG)$ is the total number of faces. A \emph{face of colors $(0c)$}, for $c\in \{1,\dotsc, D\}$, is a cycle whose edges have the colors 0 and $c$ only. One has the partition $F(\cG) = \sum_{c=1}^D F_{0c}(\cG)$.

\section{Enhanced quartic models} \label{sec:quartic}

\begin{figure}
\includegraphics[scale=.4]{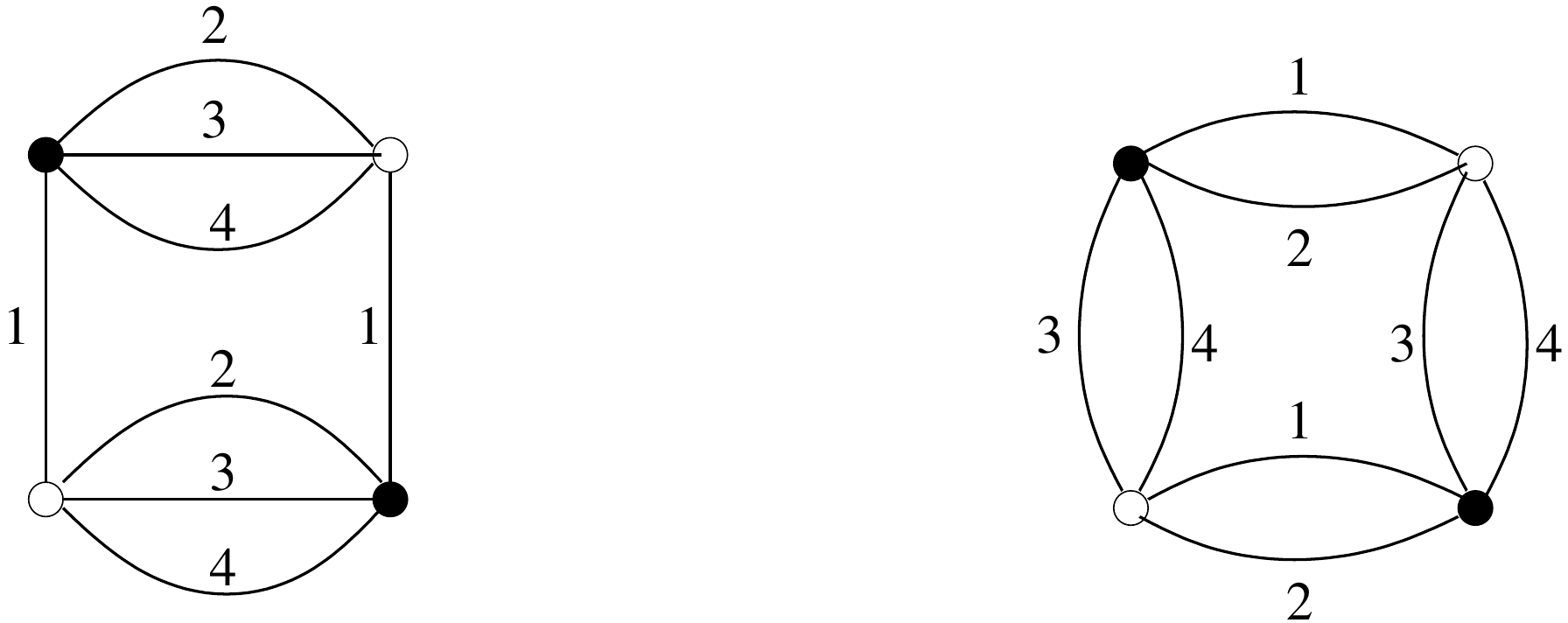}
\caption{The quartic invariants at rank 4: there are four types of melonic graphs (one of them shown left, the others being obtained by permuting the colors) and three types of necklaces (one of them shown right, the others being obtained by color permutations).} \label{order4}
\end{figure}

At rank 3, the three quartic (two black and two white vertices), connected bubbles are all melonic (and the associated model, restricted to a single coupling, i.e. invariant under permutations of the colors, was built non-perturbatively in \cite{GurauT4}). The first non-trivial maximally rescaled tensor model with quartic interactions has to be looked for at $D=4$. At that rank there are seven quartic bubbles, four of which are melonic and three non-melonic, as depicted in Figure \ref{order4}. We call the non-melonic bubbles \emph{necklaces}.

\subsection{Rank 4 Quartic Tensor Models}

We simplify slightly the notations of \cite{Delepouve:2014bma}, to adapt to the rank 4 case. The normalized Gaussian measure associated  to the unique quadratic invariant \eqref{uniquequadra} is 
\be \label{GaussianMeasure}
d\mu_{0}  = \Bigl( \prod_n \frac{ d \overline{T}_{{n}}  dT_n }{2 \imath \pi} \Bigr) \,
 e^{ - N^3 \overline{{\bf T}} \cdot_{\cD  }  {\bf T} }.
\ee

A connected quartic $U(N)^4$-invariant polynomial $B_{\mathcal{C}}$ is specified by a partition of $\cD=\{1,2,3,4\}$ into two non-trivial subsets $\cC \subset \cD $ and its complement $\cD \setminus \cC $. The associated invariant can be written as a trace over the indices of $\cC$ of a partial scalar product  over the indices of $\cD \setminus \cC $ or vice-versa \cite{Delepouve:2014bma}
\be
B_{\cC}( \overline{{\bf T}},{\bf T} ) = \tr_{\cC} \Big[ \left[ \overline{{\bf T}} \cdot_{\cD\setminus \cC }  {\bf T} \right] \cdot_{\cC}
 \left[\overline{{\bf T}} \cdot_{\cD\setminus \cC }  {\bf T} \right]  \Big]  = \tr_{\cD-\cC} \Big[ \left[ \overline{{\bf T}} \cdot_{\cC }  {\bf T} \right] \cdot_{\cD \setminus \cC}
 \left[\overline{{\bf T}} \cdot_{\cC }  {\bf T} \right]  \Big]  , \ee
where we denoted $\cdot_{\cC}$ the product of operators from $V^{\otimes \cC}$ to $V^{\otimes \cC}$.

There are therefore 7 quartic connected invariants at rank 4, the 4 melonic ones which correspond
to type 1-3 partitions in which one subset is a singleton (Figure \ref{order4} left hand side), so to $B_{\cC_i}$ with $\cC_i = \{i\}$, $i=1,2,3,4$ and the three non-melonic, \emph{necklace} partitions (Figure \ref{order4} right hand side)
which correspond to type 2-2 partitions, hence to $B_{\cC_{1i}}$ with $\cC_{1i} = \{1,i\}$ with $i = 2,3 ,4$.
Remark that such necklace interactions can be interpreted also as ordinary matrix traces on an $N^2$ space,
namely $V \otimes V$.
 
More precisely in components we have:
\be
B_{\cC_1}( \overline{{\bf T}},{\bf T} ) =  \sum_{n_1,\dotsc , n_4, n'_1,\dotsc , n'_4}
 \overline{T}_{n_1 n_2 n_3 n_4} {T}_{n_1 n'_2 n'_3 n'_4} \overline{T}_{n'_1 n'_2 n'_3 n'_4} {T}_{n'_1 n_2 n_3 n_4}
\ee
and three similar formulae for $B_{\cC_2}$, $B_{\cC_3}$ and $B_{\cC_4}$, replacing the special position 1 by 2, 3 and 4. Also
\be
B_{\cC_{12}}( \overline{{\bf T}},{\bf T} ) =  \sum_{n_1,\dotsc, n_4, n'_1,\dotsc, n'_4}
 \overline{T}_{n_1 n_2 n_3 n_4} {T}_{n_1 n_2 n'_3 n'_4} \overline{T}_{n'_1 n'_2 n'_3 n'_4} {T}_{n'_1 n'_2 n_3 n_4}
\ee
and two similar formulae for $B_{\cC_{13}}$ and $B_{\cC_{14}}$ replacing the special pair 12 by 13 and 14.
 
The standard general  quartic tensor model at rank 4 is then the (invariant) perturbed Gaussian measure 
\begin{equation}\label{eq:model}
d\mu_{standard} = d\mu_{0} \, e^{-N^{3}\sum_{i=1}^4 \lambda_i   B_{\cC_i}( \overline{{\bf T}},{\bf T} ) \; - N^{3}\sum_{i=2}^4 \lambda_{1i}   B_{\cC_{1i}}( \overline{{\bf T}},{\bf T} )   } \; .
\end{equation}
The standard melonic color-symmetric quartic model at rank 4 corresponds to $\lambda_1 = \lambda_2= \lambda_3= \lambda_4 =g_1$ and to $\lambda_{12} =\lambda_{13}= \lambda_{12}=0$, 
hence it has only one coupling constant $g_1$. It has been proved Borel summable uniformly in $N$ for $g_1$ in a cardioid domain in \cite{GurauT4}. The standard color-symmetric quartic model at rank 4 corresponds to $\lambda_1 = \lambda_2= \lambda_3= \lambda_4 =g_1$, and to $\lambda_{12} =\lambda_{13}= \lambda_{12}=g_2$,
hence it has two coupling constants. It has been proved Borel summable uniformly in $N$ for $g_1$ and $g_2$ in cardioid domains in \cite{Delepouve:2014bma}. 

The maximally rescaled general quartic model at rank 4 corresponds to changing in \eqref{eq:model} the scaling of the necklace terms from $N^{3}$ to $N^{4}$. This corresponds to their more natural scaling as matrix traces over some $V \otimes V$ space\footnote{This can be seen by changing ${\bf T}$ with ${\bf T}/\sqrt{N}$. Both the Gaussian term and the necklace interactions then get a global factor $N^2$, indicating a matrix model over $V\otimes V$ where $V$ has dimension $N^2$. More details are provided in the following section.}, and is also the maximal rescaling for which the $1/N$ expansion of the tensor model still exists, 
\begin{equation}\label{eq:fullmodel}
d\mu_{max} = d\mu_{0} \; e^{-N^{3}\sum_{i=1}^4 \lambda_i  B_{\cC_i}( \overline{{\bf T}},{\bf T} ) \; - N^{4}\sum_{i=2}^4 \lambda_{1i} B_{\cC_{1i}}( \overline{{\bf T}},{\bf T} )   } \; .
\end{equation}
The maximally rescaled \emph{restricted} quartic model at rank 4 (in short called from now on the \emph{restricted model}) is the same model but without the interactions $B_{\cC_{13}}$ and $B_{\cC_{14}}$:
\begin{equation}\label{eq:modelres}
d\mu_{res} = d\mu_{0}   \; e^{ - N^{3} \sum_{i=1}^4  \lambda_i B_{\cC_i}( \overline{{\bf T}},{\bf T} ) \;  -N^{4} \lambda_{12} B_{\cC_{12}}( \overline{{\bf T}},{\bf T} )   } \; .
\end{equation}

We are interested in the generating function of the moments and the cumulants $\kappa$ of the measure $d\mu$, which are defined as:
\begin{equation}\label{Zcumulants}
Z(J, \bar{J})=\int d\mu \;\; e^{ \sum_{n} T_{n} \bar{J}_{n} + \sum_{\bar n}\bar{T}_{\bar{n} }J_{\bar{n} }} \ , \quad \quad
 \kappa(T_{n_1}\bar{T}_{\bar{n}_1}...T_{n_p}\bar{T}_{\bar{n}_p})
=\frac{\partial^{(2p)} \Bigl( \ln Z(J,\bar J) \Bigr) }{\partial \bar{J}_{n_1}
\partial J_{\bar{n}_1}...\partial \bar{J}_{n_p}\partial J_{\bar{n}_p}} \Bigg{\vert}_{J =\bar J =0}.
\end{equation}

In order to compute the cumulants of $\mu$ we need to compute the logarithm of the generating function $Z(J,\bar J)$. This is where the field-theoretic notion of Feynman graphs is useful:
the function $Z(J,\bar J)$ expands as a power series in $g_1$ and $g_2$ indexed by Feynman graphs, and its logarithm expands as power series in $g_1$ and $g_2$ indexed by the same
Feynman graphs but \emph{connected}.

\subsection{Intermediate-field representation}

\begin{figure}[h] 
\centerline{\includegraphics[width=12cm,angle=0]{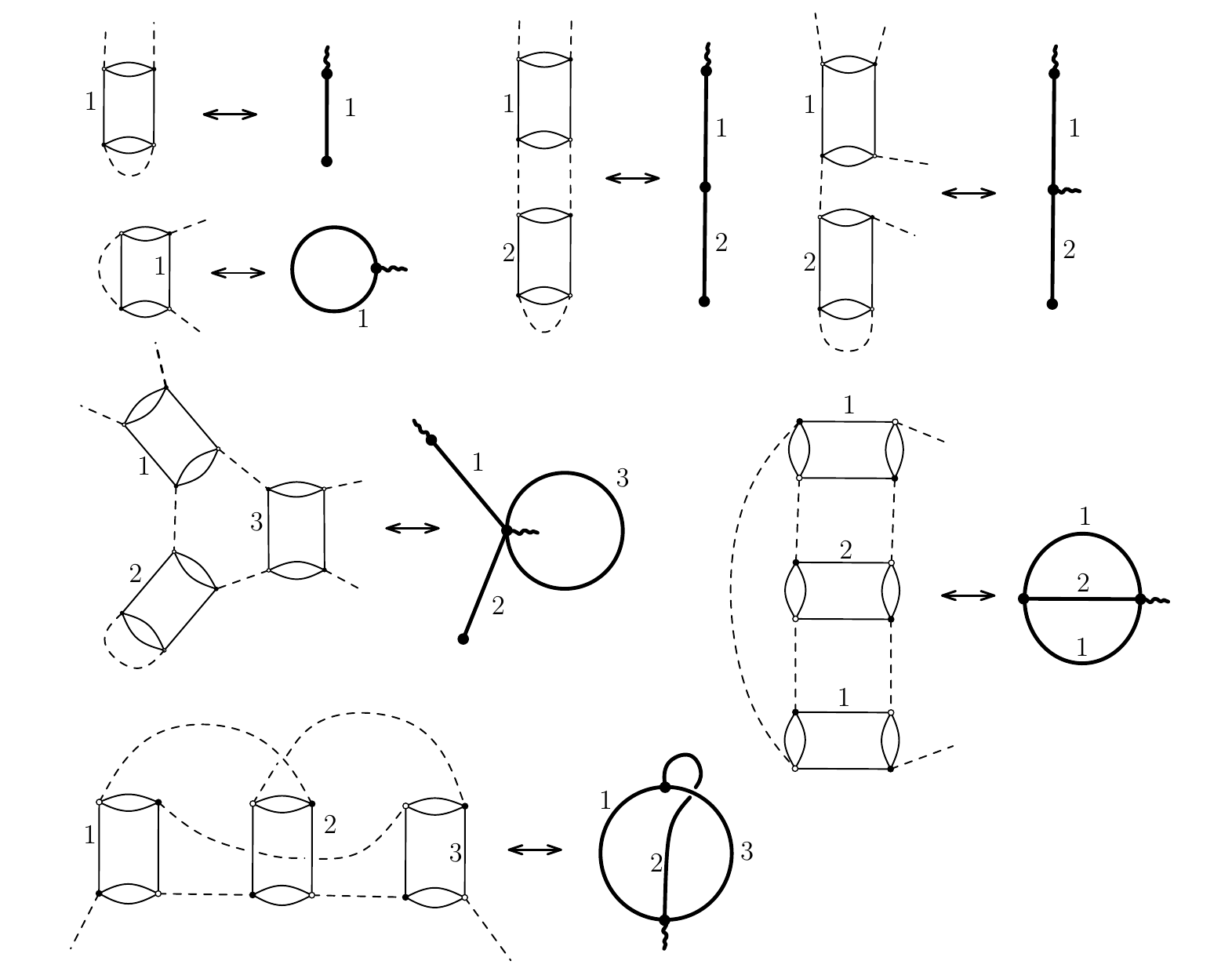}}
\caption{Intermediate Field Graphs}\label{figureIF}
\end{figure}

Quartic models are most conveniently studied in the  intermediate field (IF) representation (see Figure \ref{figureIF}). In our case it has been defined in detail in \cite{GurauT4,Delepouve:2014bma}, so that we just give an informal description sufficient for our present purpose. The graphs of this IF-representation are in fact \emph{maps}, i.e. there is a cyclic order of the edges incident to every vertex. Each IF vertex corresponds to a former cycle of 0-edges in the ordinary or O-representation, hence should be represented as a $4$-stranded loop, with strands of colors 1, 2 3 4 corresponding to the position of the tensor indices. The corners (between two incident edges) of the IF representation represent the former edges of color 0 of the O-representation. Bubbles of the type $B_{\cC_i}$ of the O-representation become \emph{monocolored} edges labeled with the color $i$, opening the single strand of color $i$ of the IF vertex. Bubbles of the type $B_{\cC_{1i}}$ in the O-representation become \emph{bicolored} edges of color type $1i$, opening simultaneously both strands of colors $1$ and $i=2, 3, 4$ of the IF vertices. Hence an IF graph of the full quartic model is a map with edges labeled by seven different color types (four monocolored possibilities and three bicolored types $1i$) and an IF graph of the restricted quartic model is a map with edges of five different labels (four monocolored or the pair 12). 

Finally external pairs of half-edges become \emph{cilia} on the vertices in the IF representation; the exact correspondence 
is quite subtle and described in detail in \cite{GurauT4,Delepouve:2014bma}, to which we refer the reader; let us simply remark that a cyclic IF vertex can bear either 0 or 1 cilium but no more.

Also rather than computing a general cumulant, we may be interested in the expectation value of a particular (generally not connected) invarant\footnote{Tensor invariants are both the interactions and the observables of a tensor model.}. The relationship between cumulants and invariants is subtle and involves Weingarten functions \cite{GurauT4,Delepouve:2014bma}. For simplicity we shall consider most of the time the simplest\footnote{The function $\log Z \vert_{J=\bar J=0}$, which expands into connected maps without cilia ($p=0)$, is apparently even simpler, but in fact the lack of a natural root slightly complicates its combinatorics. Also it requires to factorize first an overall factor $N^4$ in each vacuum map before attempting to compute the $1/N$ expansion. This $N^4$ factor is the analog of the usual volume to factor in a thermodynamic limit and corresponds to the computation of an intensive rather than extensive quantity.} function of the theory, namely the 2-point function 
\be
G_2 = \frac{\partial^{(2)} \Bigl( \ln Z(J,\bar J) \Bigr) }    
{\partial \bar{J}_{n_1}  \partial J_{n_1}    } \Bigg{\vert}_{J =\bar J =0}.
\ee 
It corresponds to the expectation value of the unique invariant of order $p=1$ defined in \eqref{uniquequadra}. It expands over
IF connected ciliated maps with a single ciliated vertex which provides a natural \emph{root} for the entire map,
and Weingarten factors are trivial. More complicated observables can be also treated by the same method.

Consider a map $\cM$ of the IF representation associated to a graph $\cG$ of the O-representation, and let us translate formula \eqref{Amplitude} in the IF language for the full quartic model \eqref{eq:fullmodel}. We denote $E_i$, $i=1,2,3,4$, the number of edges in $\cM$ with color $i$, and $E_{1i}$ the number of bicolored edges of color type $(1i)$, $i=2,3,4$. Moreover, the faces of color $0i$ in $\cG$ become the faces of the submaps $\cM_i$ which contain all the edges of color $i$ and color type $1i$ and only them (a vertex with no such incident edge is an isolated vertex in $\cM_i$ which has one face). We can still think of them as faces of colors $0i$ since the color $0$ labels the corners of $\cM$. Denote $F_{0i}$ the number of faces of colors $0i$, then
\begin{equation}\label{IF}
A(\cM) =  \prod_{i=1}^4 (-\lambda_i)^{E_i} \prod_{j=2}^4 (-\lambda_{1j})^{E_{1j}}\ N^{-  \Omega(\cM) } 
\end{equation}
where the exponent $\Omega(\cM)$ of $1/N$ in the amplitude for a map $\mathcal{M}$ is 
\begin{equation}
\Omega(\cM) = 3\sum_{i=1}^4 E_i + 2 \sum_{c=i}^4 E_{1i} - \sum_{i=1}^4 F_{0i} .
\end{equation}

\subsection{The $1/N$ expansion}
\label{sec:1/N}

Before moving on to the main results of this section, we prove the following lemma which will be used throughout the section. We recall that a \emph{cut-edge}, or a bridge, is an edge of a connected map whose deletion transforms a connected map into two connected maps.

\begin{lemma} \label{lemma.EdgeDeletion}
Let $\cM$ be a connected map and $e\in\cM$ an edge that is not a cut-edge. Denote $\cM_e$ the connected map obtained by removing $e$. Then,
\begin{enumerate}
\item\label{MonoCase} either $e$ is a monocolored edge and $\Omega(\cM)\geq \Omega(\cM_e)+2$,
\item\label{BiCase} or $e$ is a bicolored edge and $\Omega(\cM)\geq \Omega(\cM_e)$.
\end{enumerate}
Moreover, if instead $e\in \cM$ is a monocolored cut-edge which separates $\cM_1$ and $\cM_2$, then $\Omega(\cM) = \Omega(\cM_1) + \Omega(\cM_2) + 4$.
\end{lemma}

\prf Denote $\delta O = O(\cM_e) - O(\cM)$ for any quantity $O$ defined on $\cM_e$ and $\cM$. We first prove that if the color label of $e$ contains the color $i$ then $|\delta F_{0i}|\leq 1$. Indeed, there are at most two faces of color $0i$ passing along $e$ in $\cM$. If there are two distinct faces, they are merged together upon deleting $e$. If there is a single face along $e$, the deletion can at most split it into two.

\begin{itemize}
\item Then, if $e$ is monocolored of color $i$, $\delta\Omega = 3\,\delta E_i - \delta F_{0i}$. Therefore $\delta\Omega \leq 3\,\delta E_i + |\delta F_{0i}|$. With $\delta E_i = -1$, this implies $\delta\Omega \leq -2$.

\item If $e$ is bicolored with colors $1i$, then $\delta\Omega = 2\,\delta E_{1i} - \delta F_{01} - \delta F_{0i} \leq 2\,\delta E_{1i} + |\delta F_{01}| + |\delta F_{0i}| = -2 + 1 + 1 = 0$.
\end{itemize}

If $e$ is monocolored of color $i$ and separates $\cM_1$ and $\cM_2$, then removing $e$ increases $F_{0i}$ by 1 and decreases $E_i$ by 1, all other quantities being unchanged. Therefore $\delta \Omega = -\delta F_{0i} + 3 \delta E_i = -4$ and $\Omega(\cM_e) = \Omega(\cM_1) + \Omega(\cM_2)$.
\qed

Establishing the existence of the $1/N$ expansion amounts to proving that $\Omega (\cM)$ is positive for all maps, up to an overall $N^4$ for vacuum maps. 

\begin{theorem} \label{Ntheor}For any vacuum connected map $\Omega(\cM) \ge -4$; for any two-point connected map $\Omega(\cM) \ge 0$ and for any connected map for a higher order cumulant 
$\Omega(\cM) \ge 2$. Moreover, for any vacuum tree $\Omega(\cT) = -4$; for any 2-point tree $\Omega(\cT) = 0$.
\end{theorem}
The same theorem has been already established (e.g. in \cite{Delepouve:2014bma}) for the standard scaling 
$\Omega'(\cM) = 3(\sum_{i=1}^4 E_i + \sum_{c=i}^4 E_{1i} ) - \sum_{i=1}^4 F_{0i} $. But this standard scaling 
is obviously strictly greater than $\Omega (\cM)$ if $\sum_{c=i}^4 E_{1i}  >0$. Hence Theorem \ref{Ntheor} is a non-trivial new result. The fact that any cumulant of order higher than 2 is suppressed as $N\to \infty$ is similar to the standard asymptotic freeness in matrix or tensor models \cite{universality}.

\prf We first show that for any closed connected map $\cM$, and any spanning tree $\cT \subset \cM$, $\Omega(\cM)\geq \Omega(\cT)$. Then, we show that $\Omega$ is constant on trees.

Let $\cT$ be a spanning tree in $\cM$. We delete the edges in the complement and track the variations of $\Omega$ in the process using Lemma \ref{lemma.EdgeDeletion}. Order the edges in $\cM\setminus \cT$ arbitrarily from $1$ to $S=|\cM\setminus\cT|$ and define the finite sequence of maps $(\cM^{(s)})_{s=0,\dotsc,S}$ with $\cM^{(0)}=\cM$ and for $s=1,\dotsc,S$, $\cM^{(s)}$ as $\cM^{(s-1)}$ with the edge $e_s$ removed (and $\cM^{(S)} = \cT$). None of the edges in $\cM\setminus \cT$ are cut-edges, so that all the maps $\cM^{(s)}$ are connected. Moreover, Lemma \ref{lemma.EdgeDeletion} applies at each step. If $e_s$ is monocolored $\Omega(\cM^{(s)}) + 2\leq \Omega(\cM^{(s-1)})$, and if $e_s$ is bicolored, $\Omega(\cM^{(s)}) \leq \Omega(\cM^{(s-1)})$. Clearly this leads to $\Omega(\cT)\leq \Omega(\cM)$.

Notice that the bound obtained upon a monocolored deletion is strict, $\Omega(\cM^{(s)})<\Omega(\cM^{(s-1)})$, while it is soft for bicolored deletions. This reflects the fact that in the strictly monocolored case, i.e. the melonic case, only trees can dominate the large $N$ limit, while the bicolored edges may (and actually do) change that result.

Now we prove that all vacuum trees scale the same, then get a bound for trees with a fixed number of cilia.
\begin{itemize}
\item If $\cM$ is the vacuum map reduced to a single vertex, it has amplitude $N^4$ (one face for each submap $\cM_i$ of color $i$). Adding a monocolored edge of type $i$ will add one vertex, hence three new faces (those for $j\neq i$) and a factor $N^{-3}$, hence is neutral in $N$. The same thing holds for adding a bicolored edge of type $1i$: it adds two new faces and a factor $N^{-2}$, hence is neutral in $N$. Then, by induction, every vacuum map that is a tree scales like $N^4$.

\item Every 2-point tree map is obtained by adding a single cilium to a vacuum tree; this cilium opens exactly 4 strands, hence faces, hence  every 2-point map that is a tree scales at $\mathcal{O}(1)$. Finally a tree for any cumulant of higher order is obtained by adding at least one more cilium to a 2-point tree. Considering the unique path in the tree between the two cilia, we see that the new cilium opens at least two new faces, that minimum being realized when the path is made of bicolored edges all of the same type. Therefore
every higher-than-2-point tree map scales at most as $N^{-2}$.
\end{itemize}
\qed

\subsection{Characterizing LO maps}

In this subsection we prove the results for the vacuum case only, the 2-point case being similar. 

\begin{definition}
A map saturating the bound of the theorem is said to be \emph{LO} (which stands for \emph{leading order}). For simplicity we shall again
restrict to vacuum or 2-point maps and define a LO connected (vacuum or 2-point) map
as a map saturating the bound, hence with $\Omega (\cM) =-4$ (vacuum case) or $\Omega (\cM )=0$ (2-point case). Trees are LO.
\end{definition}

\begin{lemma} \label{corsub}
If a connected map $\cM$ is LO, all its connected submaps $\cM' \subset \cM$ must also be $LO$.
\end{lemma}

\prf 
Suppose $\cM' \subset \cM$
is vacuum and \emph{not} LO. We can pick a spanning tree $\cT \subset \cM$ whose restriction $\cT'$ to $\cM'$ is 
a spanning tree of $\cM'$. Then $ \Omega(\cM) = \Omega(\cT) = \Omega(\cT') < \Omega(\cM')$.
\begin{itemize}
\item Let $\tilde{\cM} = \cM' \cup (\cT\setminus\cT')$ be the map obtained from $\cM$ by deleting the edges which are neither in $\cT$ nor in $\cM'$. Performing those deletions one after another one, one gets a sequence of maps which are all connected and by Lemma \ref{lemma.EdgeDeletion}, $\Omega(\cM)\geq \Omega(\tilde{\cM})$. (Notice that here it is not necessary to pay attention to whether the deleted edges are monocolored or bicolored.)

\item Deleting the edges of $\cT\setminus\cT'$ from $\tilde{\cM}$ then does not change the value of $\Omega$. Indeed, $\tilde{\cM}\setminus\cM' = \cT\setminus\cT'$ is a forest of trees, each of which can be rooted at the vertex which meets with $\cT'$ (there is only one such vertex; otherwise, there would be a path from one to another going through $\cT'$ and another path along $\cT\setminus\cT'$ which is impossible as $\cT'\subset \cT$). It is then easy to check that deleting a leaf together with its incident edge leaves $\Omega$ invariant. By induction from the leaves to the roots of $\cT\setminus \cT'$, it comes that $\Omega(\cM) \geq \Omega(\tilde{\cM}) = \Omega(\cM')$. 
\end{itemize} 
\qed

We now want to characterize the LO maps. The above proposition implies that:
\begin{prop}
For any connected LO map,
\begin{enumerate}
\item\label{TreeCase} Any connected submap made of monocolored edges is a tree.
\item\label{PlanarCase} For all $i\neq 1$, any connected submap formed by bicolored edges with colors $1i$ is planar.
\item\label{Planar1Case} The connected submaps made of bicolored edges only are planar.
\end{enumerate}
\end{prop}

\prf By Lemma \ref{corsub}, it is sufficient to prove that the connected LO maps with only monocolored edges are trees, and that the LO maps with only bicolored edges are planar.
\begin{itemize}
\item {\it Item \ref{TreeCase}}\;\; The LO maps formed by monocolored edges correspond to the melonic sector of the quartic tensor models and consist of trees in the IF-representation \cite{GurauT4}.

\item {\it Item \ref{PlanarCase}}\;\;  It is a direct consequence of Lemma \ref{corsub} applied to Item \ref{Planar1Case} of the proposition. However, we can offer a simple and direct proof, so that we can use this result to prove Item \ref{Planar1Case}. We set $i=2$ for definiteness, and denote $\cM_{12}$ the connected map. As an ordinary map (that is, forgetting the edge colors), we denote the number of faces, edges and vertices $F_{12}, E_{12}$ and $V_{12}$. Notice that its faces with colors 01 and those with colors 02 coincide, which implies $F_{01} = F_{02} = F_{12}$. Moreover, each vertex contributes to one face with colors 03 and one with colors 04. Therefore $F_{03} = F_{04} = V_{12}$. Consequently, the function $\Omega$ evaluated on such a connected submap reduces to
\begin{equation*}
\Omega(\cM_{12}) = 2\,E_{12} - F_{01} - F_{02} - F_{03} - F_{04} = -2(F_{12} - E_{12} + V_{12}) = 4g_{12} - 4,
\end{equation*}
where the genus $g_{12}\geq 0$ of $\cM_{12}$ has been introduced. Therefore, minimizing $\Omega(\cM_{12})$ is equivalent to the map being planar, $g_{12}=0$.

\item {\it Item \ref{Planar1Case}}\;\;  Denote $\cM$ a connected map made of bicolored edges only. When seen as an ordinary combinatorial map, $\cM$ has $V$ vertices, $E=\sum_{j=2}^4 E_{1j}$ edges and $F$ faces and we want to show that its genus $g$ vanishes whenever $\cM$ is a LO map. Notice that the $F$ faces coincide with the $F_{01}$ faces of colors $01$. Therefore Euler's formula for the genus can be used to write the evaluation of $\Omega$ on $\cM$,
\begin{equation}
\Omega(\cM) = 2 E - F - F_{02} - F_{03} - F_{04} = 2g-2 + V + \sum_{j=2}^4 \bigl(E_{1j} - F_{0j}\bigr).
\end{equation}
Then, let $j\in\{2,3,4\}$ and consider the submap $\cM_j$ made of edges of color type $1j$. From the above Item \ref{PlanarCase}, for each connected component of $\cM_j$ Euler's formula holds with trivial genus. We denote $\rho_j$ the number of connected components. A vertex of $\cM$ which is not incident to an edge of type $1j$ is counted as a connected component with vanishing genus -- it has a single vertex and a single face (we have to count those isolated vertices since such a vertex carries a face of color $0j$ and these are the ones we want to count). Therefore, the total number of vertices of $\cM_j$ is $V$ and
\begin{equation}
E_{1j} - F_{0j} = -2\,\rho_j + V \qquad \Rightarrow \qquad \Omega(\cM) = 2g-2 + 2 \Bigl(2V - \sum_{j=2}^4 \rho_j\Bigr).
\end{equation}
Finally we show that $\sum_{j=2}^4 \rho_j \leq 2V + 1$. This is proved by induction on $V$. It is true for $V=1$ since then $\rho_j=1$ for $j=2, 3, 4$. Assume the bound holds for any connected maps over $V$ vertices and let $\cM$ be a connected map over $V+1$ vertices. Let $v$ be an arbitrary vertex of $\cM$ and let $\cM'$ be the map obtained by removing $v$ and all its incident edges from $\cM$. We denote $\rho_{\cM} = \sum_{j=2}^4 \rho_j$ for $\cM$ and $\rho_{\cM'}$ the same quantity for $\cM'$. There is at least one edge connecting $v$ to $\cM'$. If there is exactly one, say of type 12, then $v$ (and its possible incident loop edges) counts as a connected component for $\cM_{13}$ and $\cM_{14}$, and $\rho_{\cM} = \rho_{\cM'} + 2$. If there are more than one, then it might be that $v$ belongs to connected components of several color types which intersect $\cM'$, which implies $\rho_{\cM} \leq \rho_{\cM'} + 2$. The induction hypothesis on $\cM'$, i.e. $\rho_{\cM'}\leq 2V + 1$, then leads to $\rho_{\cM} \leq 2(V+1) + 1$, as desired.

Therefore, the bound $\sum_{j=2}^4 \rho_j \leq 2V + 1$ implies that
\begin{equation}
\Omega(\cM) \geq 2g - 4,
\end{equation}
from which is it concluded that a map of genus $g\geq 1$ cannot be a LO map.
\end{itemize}
\qed

We also want to analyze how the various monocolored and bicolored connected submaps are connected to one another in a LO map. First, we show a stronger statement than just saying that monocolored edges form forests.

\begin{prop}
All monocolored edges of a LO map are cut-edges.
\end{prop}

This can be seen as a refinement of Lemma \ref{corsub} in the case where the submap $\cM'$ is formed of monocolored edges, for which the \emph{strict} bound of the case \ref{MonoCase} of Lemma \ref{lemma.EdgeDeletion} applies.

\prf
Let $\cM$ be a (connected) LO map and assume that $e$ is monocolored and \emph{not} a cut-edge. Then its deletion results in a map $\cM'$ which remains connected. Moreover, from the case \ref{MonoCase} of Lemma \ref{lemma.EdgeDeletion}, $\Omega(\cM') < \Omega(\cM)$. This is a contradiction. \qed


\subsubsection{LO Maps, Restricted quartic model}

In the restricted case, the maps are made of monocolored edges and bicolored edges of color type $12$ only. We introduce some notations and definitions.
\begin{definition} 
Let $\cM$ be a vacuum, connected map.
\begin{itemize}
\item Removing all monocolored edges, we get a map whose connected components are submaps made of bicolored edges only (including the trivial submap made of an isolated vertex). Let $\cM_{12}$ be the set of those connected components.
\item Denote $\cE$ the set of monocolored edges. Notice that the vertices incident to a monocolored edge belong to at least one element of $\cM_{12}$, by definition.
\item Let $G_\cM$ be the graph whose vertices are the elements of $\cM_{12}$ and whose edge set is $\cE$. 
\end{itemize}
\end{definition}

\begin{theorem} \label{thm:RestrictedLO}
The vacuum, connected LO maps are the maps $\cM$ such that $\cM_{12}$ consists of planar maps of color type 12 and whose monocolored edges are cut-edges. Moreover, the graph $G_{\cM}$ is a tree.
\end{theorem}

\prf 
We already know that if $\cM$ is LO, then $\cM_{12}$ consists of planar components and monocolored edges are cut-edges. Therefore, we only have to prove that those two constraints are sufficient. Let $\cM$ be such a map.

%

Let $e$ be a monocolored edge in $\cM$. It gives rise to a unique edge in $G_\cM$ and the other way around. Therefore we identify both edges, denoted $e$. The fact that it is a cut-edge in $\cM$ implies that it also is a cut-edge in $G_\cM$. Since this holds for all edges of $G_{\cM}$, it means that it is a tree.

It remains to show that the degree of $\cM$ is $\Omega(\cM) = -4$. We will use the tree $G_{\cM}$ to that purpose. Denote $\cM_{12} = \{\cM_{12}^{(\rho)}\}_{\rho\in R}$ where $R$ is a finite set, and every $\cM_{12}^{(\rho)}$ is a connected bicolored (planar) submap of $\cM$ as well as a vertex of $G_\cM$. We equip $G_{\cM}$ with a root vertex, i.e. a distinguished $\rho^*\in R$ and a corresponding root submap $\cM_{12}^{(\rho^*)}$. That induces a partial order relation on $\cM_{12}$. $\cM_{12}^{(\rho^*)}$ is declared the largest element and its adjacent vertices are all smaller, and so on down to the leaves.

The leaves are the elements which have no elements smaller than them. In $\cM$, they correspond to submaps incident to a single monocolored edge. For every one of them, this edge is separating. Consequently, the last part of Lemma \ref{lemma.EdgeDeletion} applies, and
\begin{equation} \label{CuttingPlanar}
\Omega(\cM^{(1)}) = \Omega(\cM) - \Omega(\cM_{12}^{(\rho)}) - 4.
\end{equation}
Here $\cM_{12}^{(\rho)}$ is the submap corresponding to the leaf of $G_{\cM}$ under consideration, which is incident to the edge $e$. $\cM^{(1)}$ is the map obtained by deleting $\cM_{12}^{(\rho)}$ and $e$ from $\cM$. Since $\cM_{12}^{(\rho)}$ is planar, $\Omega(\cM_{12}^{(\rho)}) = -4$, hence 
\begin{equation*} 
\Omega(\cM^{(1)}) = \Omega(\cM).
\end{equation*}
The graph $G_{\cM^{(1)}}$ is precisely $G_{\cM}$ with the leaf and its incident edge removed. Thus, it also is a rooted tree. One then proceeds inductively, from the leaves to the root of $G_{\cM}$. This inductively removes all the connected pieces $\cM_{12}^{(\rho)}$, $\rho\in R$, and their planarity is crucial so that the degree $\Omega$ does not increase along the removal process. 

The induction reduces $G_{\cM}$ to a single vertex (the root) which corresponds in $\cM$ to either a single vertex or a non-trivial connected planar submap of color type 12. The degree is $-4$ and this is the degree of $\cM$ too.
\qed

{\bf Remark.} An alternative to the statement ``$G_\cM$ is a tree'' (and its proof) of Theorem \ref{thm:RestrictedLO} consists in the following. Let $\cT_\cM \subset \cM$ be the submap whose vertices are the vertices of $\cM$, which contains all monocolored edges of $\cM$ as well as the bicolored edges of a spanning tree of every maximally connected submaps $\cM_{12}^{(\rho)}$ of color type 12 (not reduced to a single vertex). Then $\cT_\cM$ is a tree if and only if $\cM$ is a LO map.

Indeed, first assume that $\cM$ is LO and that $\cT_\cM$ has a cycle. If it contains at least one monocolored edge, $\cM$ cannot be LO. Therefore it must be a cycle whose edges are bicolored of color type 12. Obviously they belong to a single connected component $\cM_{12}^{(\rho)}$ for some $\rho\in R$. This is however impossible since $\cT_\cM$ only contains a spanning tree of $\cM_{12}^{(\rho)}$. As a consequence, $\cT_\cM$ is a tree if $\cM$ is LO.

Furthermore, assume that $\cT_\cM$ is a tree (this implies that monocolored edges are cut-edges) and that $\cM_{12}^{(\rho)}$ is planar for all $\rho$. By construction, the difference between $\cT_\cM$ and $\cM$ itself is a set of bicolored edges. Every one of them belongs to a unique maximally connected component of color type 12. Let $e\in\cM_{12}^{(\rho)}$ be such an edge and let $\cT_{12}^{(\rho)}\subset\cT_\cM$ be the spanning tree of $\cM_{12}^{(\rho)}$. Since the full component is planar and since $e$ is not a cut-edge, removing it does not change the value of $\Omega$ on $\cM_{12}^{(\rho)}$, which eventually gives $\Omega(\cM_{12}^{(\rho)}) = \Omega(\cT_{12}^{(\rho)})$ for all $\rho$. Moreover, the components of $\{\cM_{12}^{(\rho)}\}$ are connected together by monocolored edges which are cut-edges. Therefore the last statement of Lemma \ref{lemma.EdgeDeletion} applies and easily leads to $\Omega(\cM) = \Omega(\cT_\cM)$. As we know from the proof of the $1/N$ expansion, the degree $\Omega$ is $-4$ for trees, hence $\Omega(\cM) = -4$. A similar argument will be used in the full quartic model below.

\subsubsection{LO Maps, Full quartic model}

Consider a vacuum LO map $\cM$ in the full quartic model. We know that monocolored edges are cut-edges and that the bicolored components are planar. It remains to characterize the way the bicolored planar components can be attached to one another. Indeed, in contrast with the restricted quartic case, there are three types of bicolored edges and not all planar maps with bicolored edges are LO.

\begin{definition}
Let $\cM$ be a vacuum, connected map.
\begin{itemize}
\item Remove all edges but those of color type $12$ (respectively $13$ and $14$) as well as the isolated vertices this creates,  and denote the connected components thus obtained $\cM_{12}^{(1)},\dotsc, \cM_{12}^{(R_2)}$ (respectively $\cM_{13}^{(1)}, \dotsc, \cM_{13}^{(R_3)}$ and $\cM_{14}^{(1)}, \dotsc, \cM_{14}^{R_4)}$). 
\item Pick up a spanning tree $\cT_{1i}^{(\rho_i)}$ for each $i=2,3,4$ and $\rho_i = 1,\dotsc, R_i$. Let $\cT_\cM\subset\cM$ be the submap which contains all those spanning trees as well as all monocolored edges.
\end{itemize}
\end{definition}

\begin{prop} \label{prop:FullQuartic}
Let $\cM$ be a vacuum, connected, LO map of the full quartic model. Then $\cT_\cM$ is a tree.
\end{prop}

\prf $\cT_\cM$ obviously contains all the vertices of $\cM$ and is a connected map. Let us assume that $\cM$ is LO while $\cT_\cM$ has a cycle. This cycle cannot contain a monocolored edge. It is therefore assumed to be made of bicolored edges only. Furthermore, those bicolored edges cannot be all of the same color type. Indeed, if that were the case, then they would all belong to a single connected component $\cM_{1i}^{(\rho_i)}$ (by definition of the latter and because the cycle is connected), but this is impossible as only a spanning tree of $\cM_{1i}^{(\rho_i)}$ is part of $\cT_\cM$.

The cycle, say $\ell$, thus has at least two bicolored edges of two different color types. As a vacuum map itself, it is easy to check that its degree takes a non-LO value, $\Omega(\ell)\geq -2$. According to Lemma \ref{corsub}, it cannot be a submap of a LO map, which is a contradiction.
\qed

Proposition \ref{prop:FullQuartic} puts explicit restrictions on the gluing of the planar components in the large $N$ limit. In particular, $\cM_{1i}^{(\rho_i)}$ and $\cM_{1j}^{(\rho_j)}$ can at most share one vertex. If not, there would be a path in $\cT_{1i}^{(\rho_i)}$ and another path in $\cT_{1j}^{(\rho_j)}$ joining the same two vertices, thus creating a cycle in $\cT_\cM$.

Furthermore, there cannot be ``closed chains'' of planar components of different color types. If $\cM_{1i}^{(\rho_i)}$ and $\cM_{1j}^{(\rho_j)}$ are not incident to one another, then they can be connected through either monocolored (cut-)edges, or other connected submaps $\cM_{1k}^{(\rho_k)}$ whose removal disconnects $\cM$.

\begin{figure} 
\includegraphics[scale=.35]{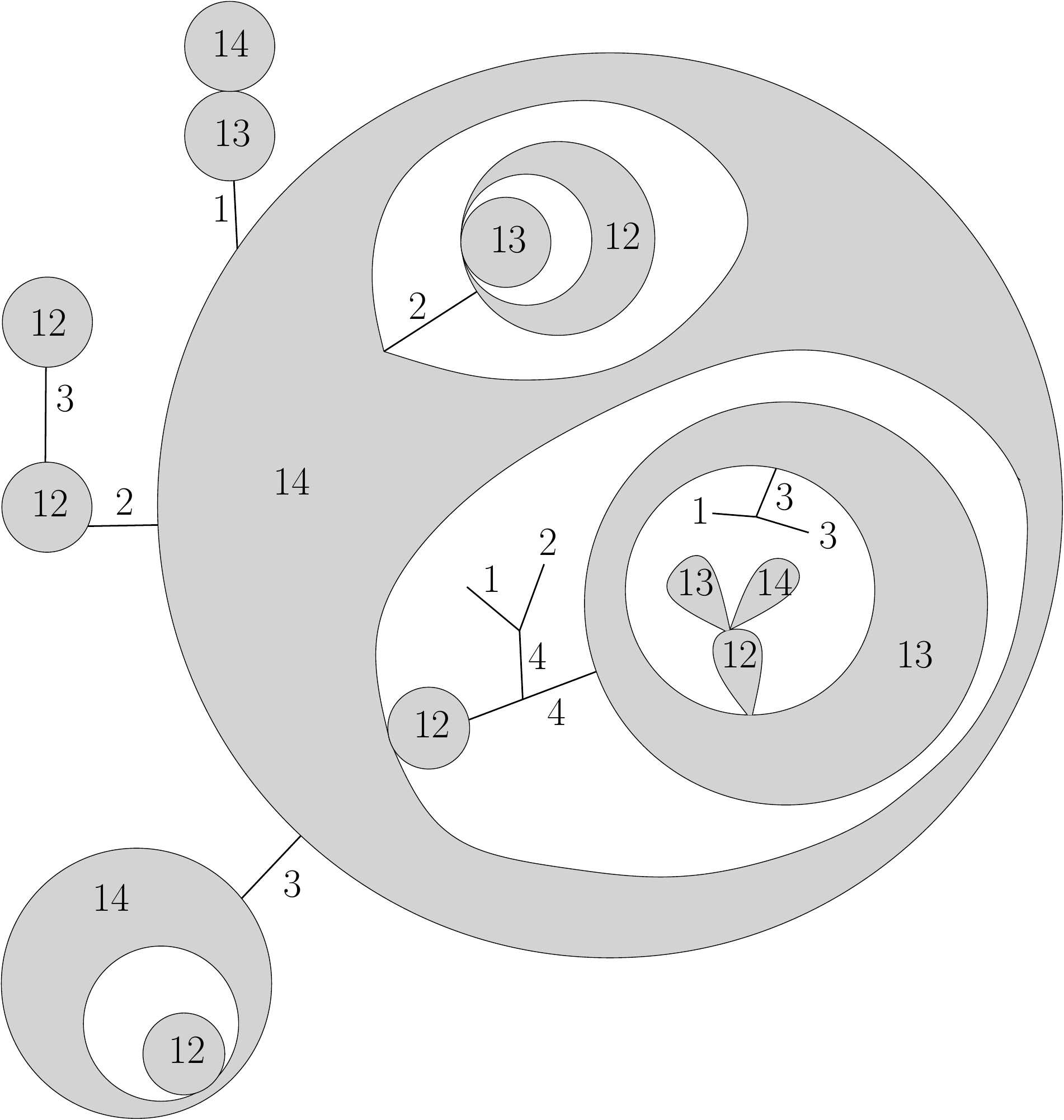}
\caption{\label{fig:multitreedisk} This represents the structure of the LO maps in the full quartic model. The grey areas are connected components of given color types. A bicolored connected component can be attached to another one on a single vertex, without forming cycles of such components.}
\end{figure}

Figure \ref{fig:multitreedisk} therefore shows the structure of the LO maps. They are planar, and made of trees of monocolored edges
which connect bicolored connected objects. The latter can touch one another at a single vertex at most and do not form closed chains, thus displaying a ``cactus'' structure.

\section{Trees of necklaces and trees of disks} \label{sec:NecklaceTrees}

\subsection{Trees of necklaces as generalizations of the quartic interactions}

Results established in the frame of a model with fixed interactions are expected to hold beyond that particular model. This is the content of the expected universality (in the sense used in statistical mechanics): changing the microscopic details (the form of the building blocks) does not affect the critical properties. This is quite well verified in matrix models. There, it is intuitive that moving from triangulations to quadrangulations or hexangulations is imperceptible in the scaling limit (mixing various interactions also brings additional degrees of freedom which upon fine-tuning leads to multicritical behaviors). We know this is also true in tensor models equipped with their standard scaling.

Even off-criticality, one can draw interesting conclusions about generic models from the study of a single one. In the 2D case, for instance, this is based on the observation that hexagons can be obtained by merging two squares. We will apply this technique to the restricted case so as to extend the range of applicability of our results to a family of models. The restricted case indeed has a quite natural generalization:
\begin{itemize}
\item change the quartic melonic interactions into arbitrary melonic interactions
\begin{equation} \label{MelonicPotential}
\nu_{melonic} = \exp -N^3 \sum_{\text{melonic $B$}} \lambda_B\ B({\bf T},\overline{{\bf T}}),
\end{equation}
\item change the quartic necklace interaction into a finite linear superposition of necklaces of the same color type but of arbitrary length
\begin{equation} \label{MatrixPotential}
\nu_{necklace(12)} = \exp -\sum_{p} \mu_p\ N^{2+p}\, B_{12}^{(p)}({\bf T}, \overline{{\bf T}}).
\end{equation}
The bubble corresponding to $B_{12}^{(p)}$ has $2p$ vertices and a single cycle made of edges alternating the colors 1 and 3.  Moreover, for each edge of color 1, there is an edge of color 2 between the same vertices. Similarly, for each edge of color 3 of the cycle, there is an edge of color 4 between the same two vertices. Note that for $p=1$ and $2$, this reproduces the Gaussian and necklace term of the measure $d\mu_{res}$.
\end{itemize}
If one turns off the necklace couplings $(\mu_p)$, this is just the ordinary tensor models with melonic interactions (which we know how to solve). If the melonic couplings are turned off on the other hand, this reduces to an ordinary matrix model. Indeed, defining $M_{AB} = \sqrt{N}\,{\bf T}_{n_1 n_2 n_3 n_4}$ whose matrix indices are pairs of tensor indices, $A=(n_1, n_2)$ and $B=(n_3, n_4)$, it is found that the necklace polynomial of degree $n$ in ${\bf T}$ is
\begin{equation*}
N^{2+p}\,B_{12}^{(p)}({\bf T},\overline{{\bf T}}) = N^2\,\tr (M M^\dagger)^p.
\end{equation*}

\begin{figure}
\includegraphics[scale=.6]{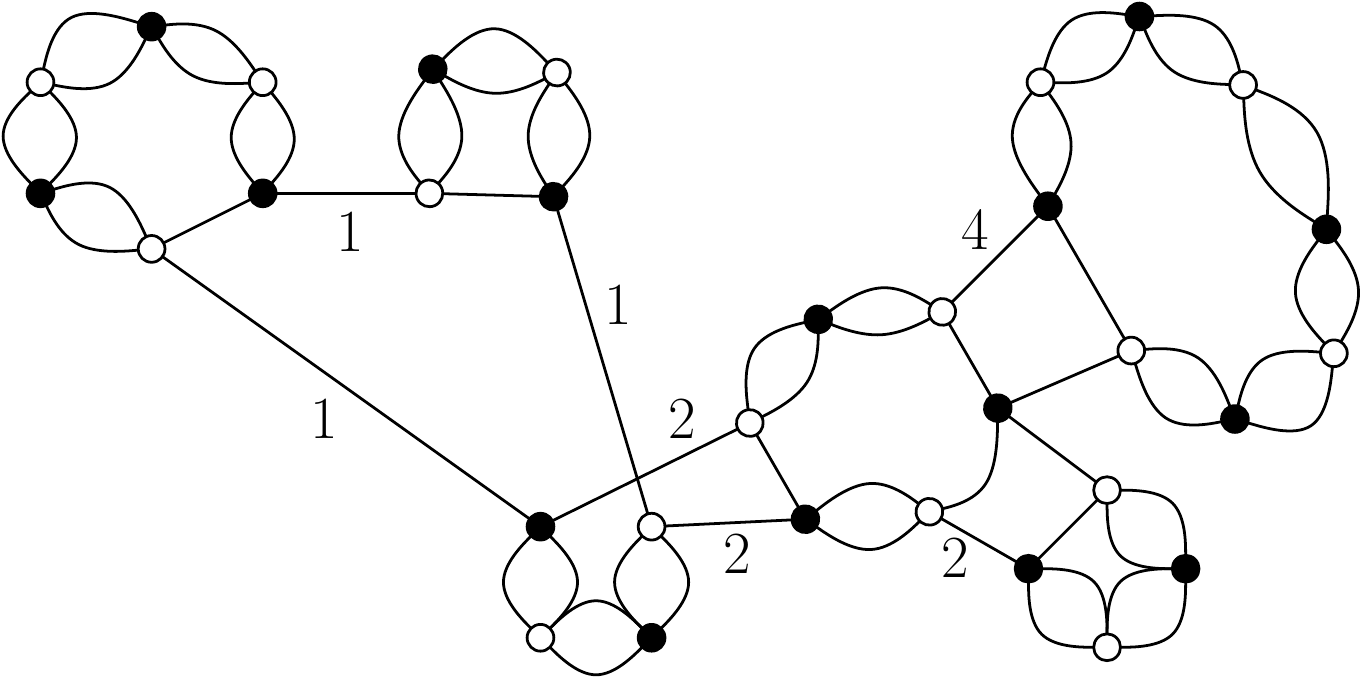}
\caption{A tree of necklaces \label{fig:TreeOfNecklaces} }
\end{figure}

Finally, we can embed the above families of melons and necklaces into a larger one, for which we give a recursive definition.
\begin{definition}
We say that a necklace (of color type 12) is open on the color $i$, for $i\in\{1, 2, 3, 4\}$, if an edge of color $i$ has been cut to form two half-edges. A \emph{tree of necklaces} of type $\{p_1,\dotsc, p_n, p_{n+1}\}$ is obtained from a tree of necklaces of type $\{p_1, \dotsc, p_n\}$ by removing any edge of color $i$ and replacing it with the necklace of size $p_{n+1}$ open on an edge of color $i$ (and preserving bipartiteness). We call this process the \emph{insertion} of a necklace (see Figure \ref{fig:TreeOfNecklaces}). 
\end{definition}

The tree of necklaces of type $\{p_1\}$ is simply the necklace of size $p_1$. The insertion of a necklace of size $p=1$ simply is a melonic insertion. Therefore we indeed reproduce all the melonic polynomials and all the necklaces of color type 12. Notice that the data $\{p_1,\dotsc, p_n\}$ does not capture the full structure of the observable. It only records the sizes of the necklaces which are inserted one after the other one. This will be sufficient for the enumeration of the LO contributions.

Let us denote a generic tree of necklaces by $\cL$. If it is of type $\{p_1, \dotsc, p_n\}$, we set
\begin{equation} \label{OmegaTreesOfNecklaces}
\omega(\cL) = \sum_{k=1}^n (2+p_k) - 3(n-1) = 3 - n + \sum_{k=1}^n p_k.
\end{equation}
This is the enhancement trees of necklaces require to contribute at large $N$\footnote{Notice that for each $k$, $2 + p_k$ is the exponent of $N$ expected for the necklace of size $p_k$. The reason for the term $-3(n-1)$ will appear in the proof of Lemma \ref{lemma:Reduction}. Let just say that those observables will be represented as trees within the quartic model with $(n-1)$ monocolored edges, each of them being supposed to come with a factor $N^{-3}$.}. The class of model we will analyze is characterized by the measure
\begin{equation} \label{NecklacesModel}
d\mu({\bf T}, \overline{{\bf T}}) = \exp \Bigl(-\sum_{\cL} N^{\omega(\cL)}\, t_\cL\,B_\cL({\bf T}, \overline{{\bf T}}) \Bigr)\ d\mu_0({\bf T}, \overline{{\bf T}}).
\end{equation}
We recall that $d\mu_0$ is the Gaussian measure \eqref{GaussianMeasure}. The sum in the exponential is over a finite number of trees of necklaces.

\begin{theorem} \label{thm:Factorization}
Let us denote the expectation of the necklace of size $p$ as
\begin{equation}
C_p = \frac{N^{2+p}}{N^4} \left\langle  B_{12}^{(p)}({\bf T}, \overline{{\bf T}}) \right\rangle = \frac{N^{2+p}}{N^4} \frac{\int d\mu({\bf T}, \overline{{\bf T}}) B_{12}^{(p)}({\bf T}, \overline{{\bf T}})}{\int d\mu({\bf T}, \overline{{\bf T}})}.
\end{equation}
Then the expectation of any tree of necklaces $\cL_{\{p_1, \dotsc, p_n\}}$ factorizes in the large $N$ limit like
\begin{equation}
\frac{N^{\omega(\cL_{\{p_1, \dotsc, p_n\}})}}{N^4} \left\langle \cL_{\{p_1, \dotsc, p_n\}}({\bf T}, \overline{{\bf T}}) \right\rangle = \prod_{k=1}^n C_{p_k}.
\end{equation}
\end{theorem}

The strategy to prove this theorem will be to reduce it to the quartic case defined by the measure \eqref{eq:modelres}, for which the proof is simple (given the results of the section \ref{sec:quartic}). To do so, we need to introduce the notion of boundary graph.

\subsection{Reduction to the quartic case}

The reduction from a generic model to the quartic model (with the standard scaling of tensor models) has already been used in \cite{DSSD}. Here we simply extend it to enhanced trees of necklaces.

An open Feynman graph is defined here as a connected, bipartite, edge-colored graph where all vertices have incident edges of colors 1, 2, 3, 4, together with some edges of color 0 which connect some black to some white vertices. In particular, some vertices are not incident to the color 0. We call them \emph{free vertices}.

A Feynman graph which is open has closed, internal faces as well as \emph{broken faces} with alternating colors $0, i$, which go from a white free vertex to a black free vertex.

\begin{definition}
The \emph{boundary graph} $\partial \cG$ of an open Feynman graph $\cG$ with $2p$ free vertices is defined as follows. Each (black or white) free vertex $h$ gives rise to a (black or white) vertex $v_h$. An edge of color $i$ is drawn between two vertices $v_{h}, v_{g}$ in $\partial \cG$ if there is a broken face in $\cG$ which starts and ends at the free vertices $h$ and $g$. The boundary graph is bipartite, regular of degree 4, with colored edge and has $2p$ vertices, but not necessarily connected.
\end{definition}

Note that a graph whose vertices are all free is its own boundary graph. The connected boundary graphs are precisely the observables of tensor models, i.e. the bubbles representing the connected tensor invariants. We recall that the latter appear as the 0-bubbles of Feynman graphs. Therefore, it makes sense graphically to trade an open, connected subgraph of a closed, connected Feynman graph for its boundary graph if the latter is connected: the result is still a connected, 5-regular, edge-colored bipartite graph. Furthermore, by definition of the boundary graph, this operation preserves all the faces which go through the subgraph but are not restricted to it (i.e. the broken faces of the subgraph).

\begin{lemma} \label{lemma:Reduction}
Let $\cF$ and $\cF_{res}$ be the sets of vacuum Feynman graphs generated by respectively the model defined by $d\mu$ \eqref{NecklacesModel} and that defined by $d\mu_{res}$ \eqref{eq:modelres}. There is a map $Q$ which to every graph $\cG\in \cF$ associates a graph $Q(\cG)\in \cF_{res}$ such that the degrees $\Omega(\cG)$ and $\Omega(Q(\cG))$ (calculated in the appropriate models) coincide.
\end{lemma}

\prf We first build the map $Q$ in a purely graphical way. This is done by exhibiting a family of open Feynman graphs of the quartic model whose boundary graphs are the trees of necklaces.

Let us start with a single necklace of size $p$. It can be obtained as the boundary graph of a loop of $p$ quartic necklaces of color type $12$ connected by edges of color 0 (drawn as dashed edges),
\begin{equation}
\begin{array}{c} \includegraphics[scale=.5]{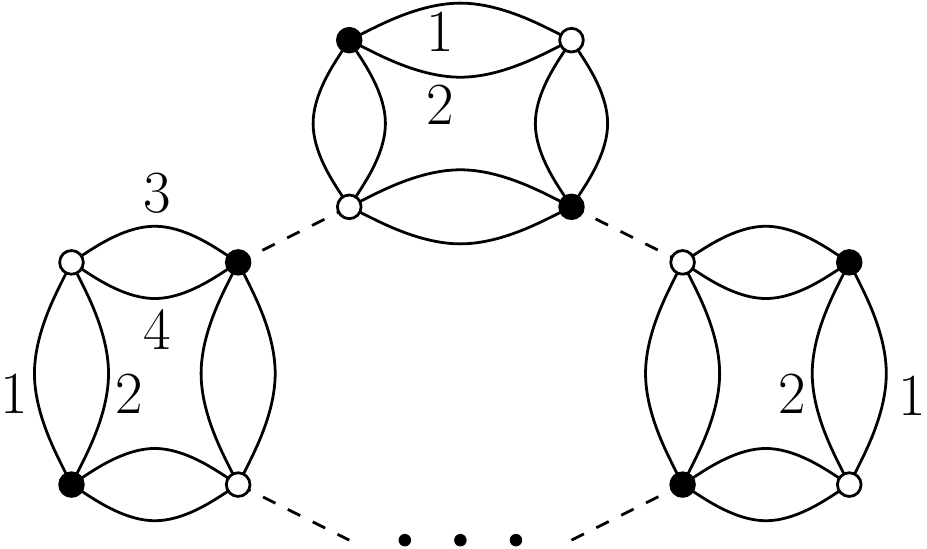} \end{array}
\end{equation}
The loop creates one closed face of colors 01 and another of colors 02 and leaves $2p$ free vertices. In the boundary graph, each quartic necklace gives rise to an edge of color 1 and an edge of color 2. The edges of colors 3 and 4 correspond to broken faces shared by couples of quartic necklaces.

Assume we have constructed an open Feynman graph $\cG_\cL$ for every tree of necklaces with $n$ insertions, such that $\partial \cG_\cL = \cL$. We consider a tree of necklaces $\cL_{n+1}$ with one more insertion. It comes from inserting an open necklace of size $p_{n+1}$ on an edge $e$ of color $i$ of a tree of necklaces $\cL_n$ of type $\{p_1, \dotsc, p_n\}$. Consider its associated open Feynman graph $\cG_{\cL_n}$. The edge $e$ is represented in $\cG_{\cL_n}$ by a broken face of colors $0i$. The white vertex to which $e$ is incident is represented in $\cG_{\cL_n}$ by a white free vertex (no edge of color 0 incident). Let us attach to this free vertex the following piece of graph
\begin{equation} \label{QuarticInsertion}
\begin{array}{c} \includegraphics[scale=.5]{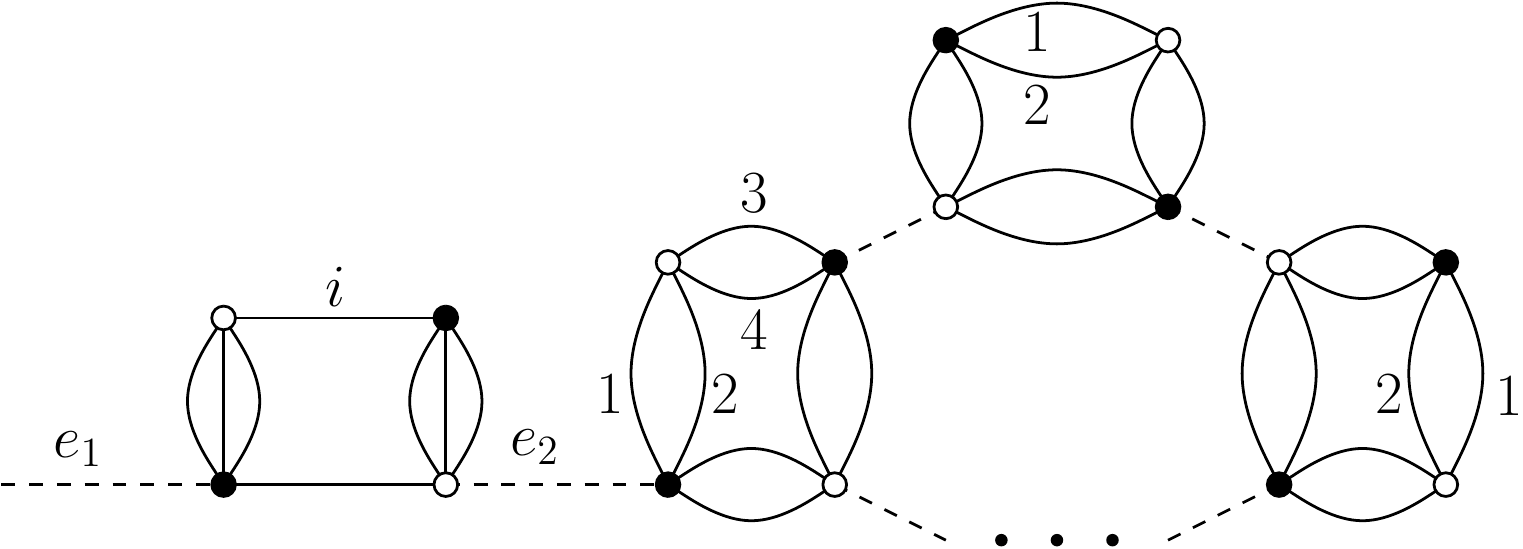} \end{array}
\end{equation}
where the loop of quartic necklaces contains $p_{n+1}$ of them. Denote $\cG_{\cL_{n+1}}$ the open graph obtained this way. We want to show that $\partial\cG_{\cL_{n+1}} = \cL_{n+1}$. Since the edges $e_1, e_2$ of color 0 are cut-edges, one can replace the subgraph $\cG_{\cL_n}$ with its boundary graph $\cL_{n}$ and also the loop of quartic necklaces in \eqref{QuarticInsertion} with its boundary graph (a necklace of size $p_{n+1}$). Denote $\tilde{\cG}_{n+1}$ the graph obtained this way, then $\partial \cG_{\cL_{n+1}} = \partial \tilde{\cG}_{n+1}$. The only portion of $\tilde{\cG}_{n+1}$ which contains non-free vertices looks like
\begin{equation}
\begin{array}{c} \includegraphics[scale=.5]{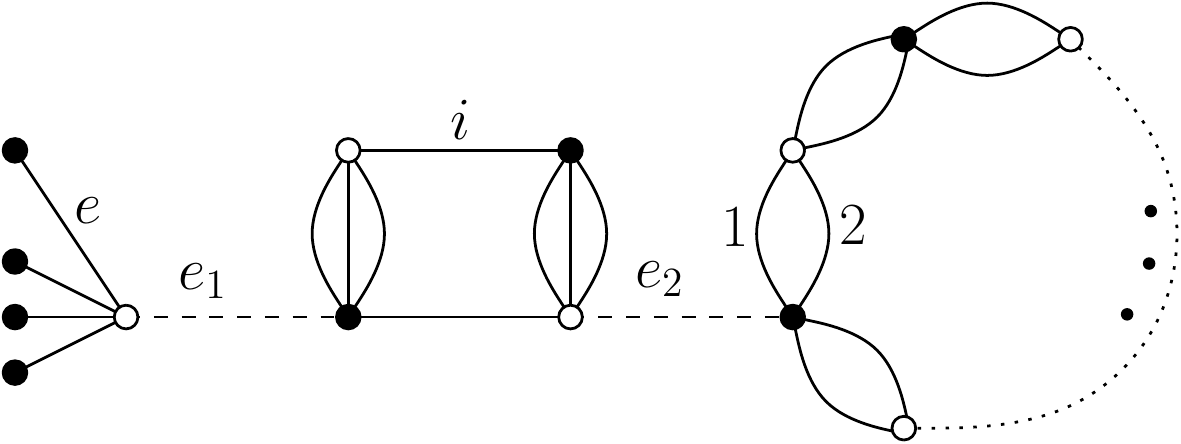} \end{array} \quad .
\end{equation}
By applying the definition of the boundary graph (and separating the cases $i=1, 2$ and $i=3, 4$), it is easy to see that it creates a necklace insertion of size $p_{n+1}$ onto $e$ in $\cL_n$, as desired.

As a consequence, given a graph $\cG\in \cF$, one can replace all its trees of necklaces $\{\cL\}$ with their associated open graphs $\{\cG_{\cL}\}$. This way, one obtains a vacuum Feynman graph $Q(\cG)\in \cF_{res}$ from the restricted quartic model.

If $\cG\in \cF$ contains $\ell(\cG)$ edges of color 0, $F_{0i}(\cG), i\in\{1, 2, 3, 4\}$, faces of colors $0i$, and a set of trees of necklaces $\{\cL\}$, it comes with weight $N^{-\Omega(\cG)}$, where
\begin{equation}
-\Omega(\cG) = \sum_{i=1}^4 F_{0i}(\cG) - 3\ell(\cG) + \sum_{\cL\subset \cG} \omega(\cL).
\end{equation}
From the definition of $\omega(\cL)$ in \eqref{OmegaTreesOfNecklaces},
\begin{equation}
\begin{aligned}
-\Omega(\cG) &= \sum_{i=1}^4 F_{0i}(\cG) - 3\ell(\cG) + \sum_{\cL\subset \cG} \sum_{k=1}^{n(\cL)} (2+p_k(\cL)) - 3(n(\cL)-1)\\
&= -3\Bigl(\ell(\cG) + \sum_{\cL\subset\cG} \bigl(2(n(\cL)-1) + \sum_{k=1}^{n(\cL)} p_k(\cL)\bigr)\Bigr) + \sum_{i=1}^4 F_{0i}(\cG) + \sum_{\cL\subset\cG} \biggl[3(n(\cL)-1) + \sum_{k=1}^{n(\cL)} \Bigl(2 + 4p_k(\cL)\Bigr)\biggr].
\end{aligned}
\end{equation}
Now we partition the above contributions in order to assign them naturally to $Q(\cG)$. We recognize 
\begin{equation}
\ell(Q(\cG)) = \ell(\cG) + \sum_{\cL\subset\cG} \Bigl(2n(\cL) - 2 + \sum_{k=1}^{n(\cL)} p_k(\cL) \Bigr)
\end{equation}
as the number of edges of color 0 of $Q(\cG)$: it still has those of $\cG$, plus those of the graphs $\cG_\cL$. In one of those graphs, we count $p_k$ edges of color 0 to form each loop of quartic necklaces whose boundary graph is a necklace of size $p_k$. If it has $n(\cL)$ necklaces, there are $2n(\cL)-2$ edges of color 0 added to connect them (like $e_1, e_2$ in \eqref{QuarticInsertion}).
We also recognize 
\begin{equation}
b_{12}(Q(\cG)) = \sum_{\cL\subset\cG} \sum_{k=1}^{n(\cL)} p_k(\cG),\qquad \text{and}\qquad b(Q(\cG)) = \sum_{\cL\subset\cG} n(\cL) - 1,
\end{equation}
the number of quartic necklaces and the number of melonic quartic bubbles of $Q(\cG)$. Finally
\begin{equation}
F_{0i}(Q(\cG)) = F_{0i}(\cG) + (\delta_{i,1} + \delta_{i,2})\sum_{\cL\subset\cG} n(\cL),
\end{equation}
is the number of faces of $Q(\cG)$ of colors $0i$, since every necklace represented as in \eqref{QuarticInsertion} has two internal faces, one of colors $01$, and one of colors $02$.

It comes
\begin{equation}
-\Omega(\cG) = -3\ell(Q(\cG)) + 4 b_{12}(Q(\cG)) + 3b(Q(\cG)) + \sum_{i=1}^4 F_{0i}(Q(\cG)) = -\Omega(Q(\cG)),
\end{equation}
as assigned in the restricted quartic model to $Q(\cG)$.
\qed

\subsection{Factorization}

{\bf Proof} of Theorem \ref{thm:Factorization}. The expectation of an observable is a sum over all the closed Feynman graphs generated by the model which contain the observable as a marked subgraph. We choose the observable to be a tree of necklaces. The graphs contributing to its expectation are closed graphs made of trees of necklaces connected through edges of color 0. We apply the map $Q$ to represent them in the quartic model and then use the intermediate-field representation. Through this process, a tree of necklace $\cL$ of type $\{p_1, \dotsc, p_n\}$ becomes an open graph $\cG_{\cL}$ which in the IF representation becomes a map $\cT_\cL$ with $n$ bicolored loops of sizes $p_1, \dotsc, p_n$ connected in a tree-like fashion (no closed circuits between those loops) via $n-1$ monocolored edges.

From the combinatorial description of the LO graphs of the restricted quartic model, it is clear that there exist graphs $\cG$ contributing to the expectation such that $Q(\cG)$ is a LO graph. From Lemma \ref{lemma:Reduction}, it thus comes that $\cG$ contributes to the LO of the expectation if and only if $Q(\cG)$ is LO in the quartic model.

Let $\cG$ be a LO graph contributing to the expectation of $\cL$ which is of type $\{p_1, \dotsc, p_n\}$. In the IF representation, $Q(\cG)$ contains $\cG_\cL$ as a marked subgraph with $n-1$ monocolored edges and $n$ bicolored loops. All those monocolored edges are cut-edges in $Q(\cG)$. Cutting them yields $n$ connected components, each one containing one of the $n$ bicolored loops as a marked subgraph. This establishes a factorization of the expectation of $\cL$ onto its necklaces,
\begin{equation} \label{Factorization}
\frac{N^{\omega(\cL_{\{p_1, \dotsc, p_n\}})}}{N^4} \left\langle \cL_{\{p_1, \dotsc, p_n\}}({\bf T}, \overline{{\bf T}}) \right\rangle = \prod_{k=1}^n \tilde{C}_{p_k}.
\end{equation}
Moreover, any set of graphs contributing to the expectations of the loops of sizes $p_k(\cL)$, $k=1, \dotsc, n$, gives rise to a LO contribution to the expectation of $\cL$ by gluing back the monocolored edges. This proves that the contribution of a necklace  is $\tilde{C}_{p_k} = C_{p_k}$ as defined in Theorem \ref{thm:Factorization}.
\qed

Although this result is simpler to prove using the IF representation, it can also be understood directly in terms of the edge-colored graphs generated by the initial model. If $\cL$ is the observable whose expectation we are interested in, let $\cL_{p_k}$ be one of its necklaces, and let $v, \bar{v} \in\cL_{p_k}$ be white and black vertices incident to edges of color $i$ which connect it to other necklaces. By construction of $\cL$, $v$ and $\bar{v}$ can be chosen such that their incident edges of color $i$ are 2-cut-edges. In $\cG$, there are also two edges of color 0 incident to $v$ and $\bar{v}$. 
\begin{itemize}
\item The factorization \eqref{Factorization} means that any path which starts at $v$ (or $\bar{v}$) and follows its incident edge of color 0 does not hit $\cL$ before it comes back to a vertex of $\cL_{p_k}$.
\item Moreover, the equality $\tilde{C}_{p_k} = C_{p_k}$ can be found by studying the number of faces of colors $0i$ which go through $\cL_{p_k}$. The unique face of colors $0i$ which crosses $v$ follows the edge of color $i$ incident to $v$ and necessarily goes along the edge of color $i$ incident to $\bar{v}$. It happens this way because the two edges of color $i$ incident to $v$ and $\bar{v}$ are 2-cut-edges, i.e. they carry a 2-point function with external edges of color $i$. Therefore, when counting the number of faces $0i$ which go through $\cL_{p_k}$, everything happens as if $v$ and $\bar{v}$ were connected by an edge of color $i$. This means that the counting can be performed directly on the isolated necklace of size $p_k$, which yields the desired equality.
\end{itemize}

\subsection{Enumeration: Large $N$ free energy and expectations}

Thanks to the combinatorial analysis of the quartic model performed in Section \ref{sec:quartic}, the problem of evaluating the expectations of trees of necklaces at large $N$ has been reduced to determining the expectations of the necklaces: this is the content of Theorem \ref{thm:Factorization}. Calculating those expectations amount to enumerating the number of Feynman graphs containing a necklace of fixed size as a marked subgraph. To do so, we will write an equation {\it \`a la Tutte}, known in our context as Schwinger-Dyson equations.

The Schwinger-Dyson equations of random tensor models have been described in \cite{BubbleAlgebra}, and the structure and large $N$ solution have been described in \cite{TreeAlgebra, LargeNSD}. We recall that an observable is a $U(N)^4$-invariant polynomial which can be represented as a connected, bipartite, $4$-colored graph. Then define the following two operations.
\begin{itemize}
\item The contraction of an observable $B$ with $2p$ vertices along the black and white vertices $\bar{v}, v \in B$, denoted $B/(\bar{v}, v)$ is obtained by connecting $v$ and $\bar{v}$ with an edge of color 0 and taking the boundary graph. Equivalently, one removes $\bar{v}$ and $v$ and reconnects the edges while respecting the colors, to get a set of disjoint observables on $2(p-1)$ vertices.

\item The composition of two observables $B, B'$ along the pair $v\in B, \bar{v}'\in B'$ is obtained by connecting $v$ to $\bar{v}'$ via an edge of color 0 and then taking the boundary graph. Equivalently, one removes $v$ from $B$ and $\bar{v}'$ from $B'$ and reconnects the edges while respecting the colors. If $B$ has $2p$ vertices and $B'$ has $2p'$ vertices, then their composition, denoted $B\star_{(\bar{v}', v)} B'$, has $2(p+p'-1)$ vertices.
\end{itemize}

Let $B$ be an observable, represented as a connected, bipartite, $4$-colored graph, with a \emph{marked white vertex} $v$. There is one Schwinger-Dyson equation for each such observable with a marked vertex. It reads
\begin{equation} \label{SD}
\sum_{\bar{v}\in B} \langle B/(\bar{v}, v) \rangle - N^3 \langle B\rangle - \sum_{\cL} N^{\omega(\cL)}\,t_\cL \sum_{\bar{v}'\in\cL} \langle B\star_{(\bar{v}', v)} \cL \rangle = 0,
\end{equation}
where $\{\cL\}$ denotes the set of trees of necklaces which defines the measure \eqref{NecklacesModel}. 

We choose $B=B_{12}^{(p)}$ the necklace of size $p$, and multiply the equation by $N^{(p-1)}/N^4$. Then
\begin{itemize}
\item If $v$ and $\bar{v}$ are separated by $k$ white vertices, then
\begin{equation}
\frac{N^{(p-1)}}{N^4}\langle B_{12}^{(p)}/(\bar{v}, v) \rangle = \frac{N^{2+(p-k-1)}}{N^4}\langle B_{12}^{(p-k-1)} \rangle\ \frac{N^{2+k}}{N^4}\langle B_{12}^{(k)} \rangle = C_{p-k-1}\,C_k,
\end{equation}
which is the same large $N$ factorization as in ordinary matrix models.
\item The middle term of the equation, which comes from the Gaussian part of the measure, reads
\begin{equation}
\frac{N^{(p-1)}}{N^4}\,N^3 \langle B_{12}^{(p)} \rangle = C_p.
\end{equation}
\item If $\cL$ is an interaction of the model, of type $\{p_1, \dotsc, p_n\}$ and $\bar{v}'\in\cL$ belongs in a necklace $\cL_{p_j}$, then the composition with $B_{12}^{(p)}$ increases its size to $p_j+p-1$, and
\begin{equation}
\begin{aligned}
\frac{N^{(p-1)+\omega(\cL_{\{p_1,\dotsc,p_n\}})}}{N^4} \langle B\star_{(\bar{v}', v)} \cL_{\{p_1,\dotsc,p_n\}} \rangle &= \frac{N^{\omega(\cL_{\{p_1, \dotsc, p_j+p-1, \dotsc, p_n\}})}}{N^4}\,\langle \cL_{\{p_1, \dotsc, p_j+p-1, \dotsc, p_n\}} \rangle\\
&= C_{p_j+p-1}\,\prod_{\substack{k=1,\dotsc,n\\ k\neq j}} C_{p_k},
\end{aligned}
\end{equation}
where the last equality makes use of Theorem \ref{thm:Factorization}.
\end{itemize}
Gathering the above pieces, equations \eqref{SD} become a system which determines the rescaled expectations $(C_1, C_2, C_3, \dotsc)$. It takes the following generic form. Let $V(x_1, x_2, x_3, \dotsc)$ be a polynomial in a finite number of variables $(x_1, x_2, x_3, \dotsc)$, and denote $\partial_p$ the derivative with respect to $x_p$. Then the Schwinger-Dyson equation has the form
\begin{equation} \label{Enumeration}
C_p = \sum_{k=0}^{p-1} C_k\,C_{p-k-1} + \sum_{j\geq 1} j\,\partial_j V(C_1, C_2, C_3, \dotsc)\ C_{j+p-1}.
\end{equation}
This equation has a clear interpretation in terms of branching planar maps (i.e. trees of disks). $C_p$ is the number of maps in the class with root vertex of degree $p$, or by duality, boundary face of degree $p$. The quadratic term corresponds, as usual for equations {\it \`a la Tutte}, to the case where the root edge is a bridge. If not, its removal extends the length of the boundary face from $p$ to $p+j-1$, which is also usual for planar maps. However, it here comes with a weight $j\partial_j V(C_1, C_2, \dotsc)$. When this is independent of the numbers $\{C_k\}$ (the linear terms of $V$), one simply recovers Tutte's equation for planar maps with bounded face degrees. All other terms of $V$ correspond to gluing ``on top'' of our map other maps with prescribed boundary face degrees. This superimposition generates a \emph{branching process} which we already analyzed at the combinatorial level in the quartic case in Section \ref{sec:quartic} in terms of trees of disks.

Equation \eqref{Enumeration} can be found in the existing literature devoted to multi-trace matrix models \cite{AlvarezBarbon, Das}. Therefore, we will only describe its critical properties without reproducing the detailed analysis. The free energy behaves like $(g-g_c)^{2-\gamma}$, where $g$ is the ``cosmological constant'' (an overall coupling, or a coupling of ordinary single-trace terms of the potential), $g_c$ is the radius of convergence of the generating function for $(C_p)_{p\geq 1}$ (it depends on all the other couplings of the model) and $\gamma$ is the \emph{entropy exponent}. The latter characterizes the proliferation of some combinatorial degrees of freedom of the maps like $n^{\gamma-3}$ where $n$ is the size of the objects. It classifies the various universality classes which can be achieved.
\begin{itemize}
\item When the planar components become critical while the branching process stays off-criticality, the latter is washed away in the scaling limit and the string susceptibility exponent of pure 2D gravity is recovered, $\gamma=-1/2$.
\item If the situation is reversed (non-critical maps, critical branching), $\gamma=1/2$, as expected for trees since the planar components become irrelevant in the scaling limit; this is the universality class of branched polymers.
\item If both are simultaneously critical (a phase reached by tuning one additional coupling to a critical value), the exponent $\gamma=1/3$ can be reached. This phase transition describes a proliferation of baby universes.
\item Tuning more couplings leads to more phases, with exponents $\gamma = p/(n+m+1)$, where $p\leq n$ and $m$ are integers. It is argued in \cite{AlvarezBarbon} that for $n>1$ the polymerization is too strong for the surface to support macroscopic loops.
\end{itemize}

\section{Analyticity of the Model}
\label{sec:analytic}

We discuss the analyticity of the 2-point function $G_2$ of the general quartic model in a cardioid domain of the complex plane, using the Loop Vertex Expansion. The proof, being similar to the one of the non-enhanced case \cite{Delepouve:2014bma}, will be discussed in less details. 

\begin{theorem}
The 2-point function $G_2$ of the maximally rescaled general quartic model with coupling constants $\lambda_\cC=|\lambda_\cC| e^{\imath \phi_\cC}$ is analytic for $|\lambda_\cC|<\frac1{56} \cos^2\frac\phi2,\ \phi={\rm max}_\cC (\phi_\cC) $.
\end{theorem}

To prove this theorem, we will first use a new series expansion of $\log Z$ in the intermediate field representation, (the Loop Vertex Expansion), then prove the absolute convergence of the series for all $\lambda$'s inside the cardioid domain.

\subsection{Loop vertex expansion}

Using a standard Hubbard-Stratonovich transformation on \eqref{eq:fullmodel}, the model can be re-written in terms of seven hermitian matrices $\sigma^\cC$, where for $i\in\{1, \dotsc, 4\}$, $\sigma^i$ is a $N\times N$ matrix, and for $i\geq2$, $\sigma^{1i}$ is a four indices tensor seen as an hermitian $N^2\times N^2$ matrix acting on pairs of indices of color $\{1,i\}$.  The measure on those matrices (known as intermediate fields) is
\bea\label{IFaction}
d\mu(\sigma)=d\mu_0 (\sigma)
e^{-{\rm Tr}_\cD {\rm ln}\left[{\bf 1}^\cD \ +\ \sum_{i=1}^4 \imath\sqrt{\frac{\lambda_i}{N^3}} \left({\bf 1}^{\cD \setminus \{i\}}\otimes \sigma^i \right)\ +\ \sum_{i=2}^4 \imath\frac{\sqrt{\lambda_{1i}}}{N} \left({\bf 1}^{\cD \setminus \{1,i\}}\otimes \sigma^{1i}\right)\right] }\ ,
\eea
where $\bf{1}^\cB$ is the identity matrix over the set of indices $\cB$ and $d\mu_0(\sigma)$ is the normalized Gaussian measure of covariance 1 over the $\sigma$ matrices.

Denoting $ A(\sigma) = \imath \sum_{\cC} \sqrt{\frac{\lambda_\cC} { N^{4-|\cC|} }} \left( {\bf 1}^{\cD \setminus \cC} \otimes \sigma^\cC \right) $ and $R(\sigma)=\left[{\bf1}^\cD + A\right]^{-1}$, the generating function of the cumulants (\ref{Zcumulants}) is
\be
Z(J,\bar J) = \int e^{N^{-3}\sum_{n,\bar n}\bar J_{\bar n} R(\sigma)_{\bar n n} J_n}d\mu(\sigma).
\ee
Hence the 2-point function can be expressed as the connected expectation value of the resolvent, i.e. 
\begin{equation}
G_2 = \frac1{Z} \int \frac{R(\sigma)}{N^3} d\mu.
\end{equation}
Moreover, it admits an expansion as a sum of trees, via the BKAR formula \cite{brydges,Abdesselam:1994ap}. For any tree $T_v$ with $v$ vertices, 
\begin{itemize}
 \item to any edge $\ell$ of $T_v$, we assign a parameter $u_\ell\in [0,1]$,
 \item for any pair of vertices $\{i,j\}$, $P_{ij}$ being the unique path joining the vertices, we define $w_{ij}={\rm min}_{\ell\in P_{ij}} u_\ell$, and we define $w_{ii}=1$, 
 \item we define a new set of $\sigma$ matrices $\{\sigma^{i\ \cC}\}$ for each vertex $i$ of $T_v$,
 \item we define the interpolated Gaussian measure $\mu_{T_v, u}$ as $\int \sigma^{i\ \cC}_{n\bar n}\sigma^{j\ \cC}_{m\bar m}\ d\mu_{T_v, u}(\sigma) = w_{ij}\delta_{n\bar m}\delta_{\bar n m}$.
\end{itemize}
Then we can express $G_2$ as a sum over plane trees $\cT_{v,\{ \cC ( \ell ) \}, i^*}$ with $v$ vertices, colored edges $\ell$ and a single ciliated vertex $i^*$. Each corner $p$ of a vertex $i$ bears a resolvent $R(\sigma^i)$ whose indices $n_{i,p}^c,\bar n_{i,p}^c$ ($c\in\{1..4\}$) are contracted along the faces of color $c$ of the tree. Therefore, the vertices have a structure of 4-stranded vertices and to an edge $\ell_{ij}$ between $i$ and $j$ corresponds the contraction
\be
 \delta^{\ell_{ij},{\mathcal{C}}(\ell_{ij})}=  
 \left( \delta_{{\bar n}^{\cD \setminus \cC }_{i,q} n^{\cD  \setminus \cC }_{i,q+1} } \right) 
    \delta_{ \bar{n}^{\cC}_{i,q} n^{\cC}_{j,p+1} }      \delta_{ \bar{n}^{\cC}_{j,p} n^{\cC}_{i,q+1} }
 \left( \delta_{\bar{n}^{\cD \setminus \cC }_{j,p} n^{\cD \setminus \cC }_{j,p+1} } \right) \; .
\ee
The ciliated vertex bears a cilium at a corner, that, for the 2-point function, is just a trivial contraction of the adjacent strands $\frac1N\delta_{n_{i,p}\bar n_{i,p+1}}$. For higher order cumulants, the strands have to be re-connected according to non-trivial permutations over cilia, and with a factor involving Weingarten functions 
\cite{GurauT4,Delepouve:2014bma,Collins1,Collins2}. The cilium acts as an edge for the ciliated vertex $i^*$, increasing its degree and number of corners by one. Thus,
\begin{multline}
\label{eq:G2LVE}
(G_2)_{\bar nn} = \frac{\delta_{\bar nn}}{N^4}
\sum_{v\geq 1}\frac1{v!}\ 
\sum_{\mathcal{T}_{v,\{{\mathcal{C}}(\ell)\}, i^*}}
\int_0^1 \left[\prod_{\ell\in \cT_v}du_{\ell}\right]\int d\mu_{\cT_v,u}(\sigma)
\sum_{n,\bar n}  \left(\prod_{i\in V(\cT_v)}\prod_{p=1}^{\mathrm{degree}(i)} R(\sigma^i)_{\bar{n}_{i,p}n_{i,p}}\right)\\
\times \left(\prod_{\ell\in \cT_v}\frac{-2\lambda_{\cC(\ell)}}{N^{4-|\cC(\ell)|}} \ \delta^{\ell,{\mathcal{C}}(\ell)}\right)\ .
\end{multline}
\begin{figure}[h!]
\includegraphics[scale=.25]{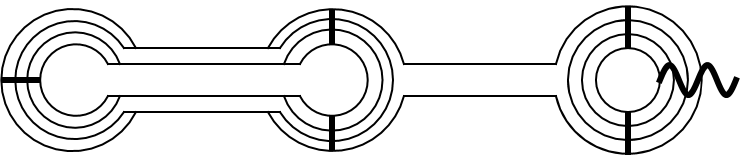}
\caption{A 4-stranded tree with a cilium. Resolvents are represented as thick bars crossing each strand of a corner. } \label{LVEtree}
\end{figure}
\subsection{Cardioid domain}

The series (\ref{eq:G2LVE}) is absolutely convergent for coupling constants $|\lambda_\cC|e^{i\phi_\cC}, |\phi_\cC|<\pi$ inside a cardioid domain $|\lambda_\cC|< \rho \cos^2\left( \frac{\phi}2\right)$, $\phi={\rm max}_\cC\ \phi_\cC$. This can be shown by bounding the contribution of each tree $\mathcal{T}_{v,\{{\mathcal{C}}(\ell)\},i^*}$ in (\ref{eq:G2LVE}) using iterated Cauchy-Schwarz inequalities \cite{Delepouve:2014bma}. Writing
\bea
C(\mathcal{T}_{v,\{{\mathcal{C}}(\ell)\},i^*})=\left(\prod_{i\in V(\cT_v)}\prod_{p=1}^{\mathrm{d}(i)} R(\sigma^i)_{\bar{n}_{i,p}n_{i,p}}\right)
\prod_{\ell\in \cT_v} \delta^{\ell,{\mathcal{C}}(\ell)} ,
\eea
a colored plane tree $\cT$ made of 4-stranded vertices but with no resolvent is structurally an intermediate field Feynman graph and thus comes with the amplitude $\cA(\cT)$ as in \eqref{IF}. Resolvents can be taken care of by applying a by now rather standard technique 
\cite{Delepouve:2014bma,Magnen:2009at,Rivasseau:2014bya,Delepouve:2014hfa}. Starting at the cilium, we order the $(2v-1)$ resolvents along the clockwise contour walk around the tree, indexing them as $R_1$, $R_2$\ldots Then, we draw a line along the unique path between $R_1$ and $R_{v}$. Everything left (respectively right) of this path is called $A$ (resp. $B$), then we use the following Cauchy-Schwarz bound:
\be
C(\mathcal{T}_{v,\{{\mathcal{C}}(\ell)\},i^*})=\bra{A} R_1 \otimes R_v \otimes {\bf1}^{\otimes k}\ket{B} \leq \|R_1\|\|R_v\|\sqrt{\braket{A|A}\braket{B|B}} .
\ee
\begin{figure}[h!]
\includegraphics[scale=.2]{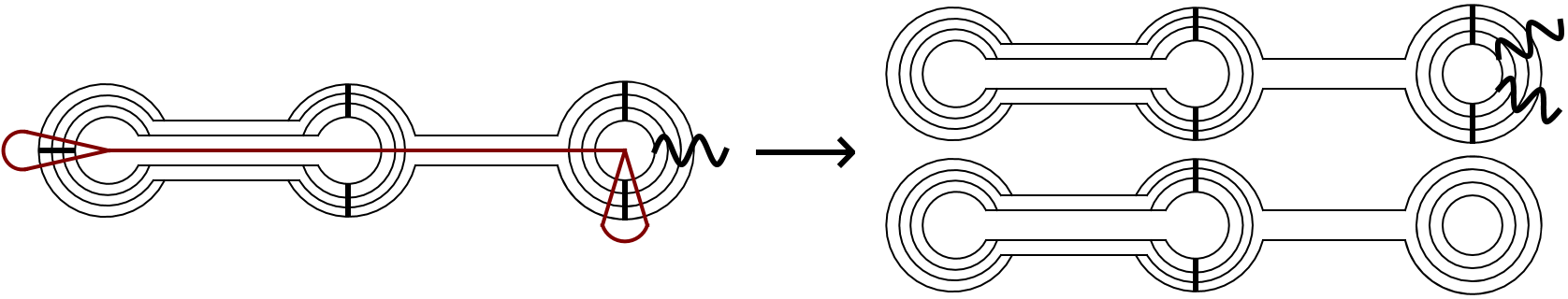}
\caption{First Cauchy-Schwarz iteration applied on the tree from Fig.\ref{LVEtree}. The cilia could have been deleted as they are merely identity operators. } \label{CauchySchw}
\end{figure}

Within the cardioid, $\|R(\sigma)\|\leq 1/\cos(\phi/2)$ with $\phi = \max_\cC\ \phi_\cC$. Moreover, the scalar product graphs $\braket{A|A}$ and $\braket{B|B}$ have the same structure of plane trees with resolvents as the original graph, except that $\braket{A|A}$ now bears $2v-4$ resolvents and $\braket{B|B}$ has $2v-2$. Repeating the same process on $\braket{A|A}$ (but using arbitrary corner as $R_1$) will give rise to graphs bearing $2v-6$ resolvents and so on. Repeating the process until no resolvents remains on any graph, we have 
\bea
C(\mathcal{T}_{v,\{{\mathcal{C}}(\ell)\},i^*}) &\leq&
\prod_{j=1}^{2v-1} \| R_j(\sigma^{i(j)})\| \prod_{\mathfrak{g}_A} N^{-\frac{F (\mathfrak{g}_A)}{2^{v-1}}}\prod_{\mathfrak{g}_B} N^{-\frac{F (\mathfrak{g}_B)}{2^{v}}}\nonumber\\
&\leq& \left(\cos\frac{\phi}2\right)^{-(2v-1)} N^{F(\cT_v^{\setminus R})},
\eea
where $F(\cG)$ is the number of faces of $\cG$, $\mathfrak{g}_A$ and $\mathfrak{g}_B$ are the graphs bearing no resolvents created from the right amount of iterations of the Cauchy-Schwarz bound on the graphs $\braket{A|A}$ and  $\braket{B|B}$ and $\cT_v^{\setminus R}$ is the tree with all resolvents removed. The second inequality arises from the conservation of the number of faces during the Cauchy-Schwarz process. Indeed, at each iteration, the number of faces is multiplied by 2.

For any tree, $F(\cT_v^{\setminus R}) = 4v - \sum_\ell \cC(\ell)$, therefore,
\bea
|(G_2)_{ nn}| &\leq& 
\sum_{v\geq 1}\frac{1}{v!}\cos\frac{\phi}2\ 
\sum_{\mathcal{T}_{v,\{{\mathcal{C}}(\ell)\}, i^*}}
\prod_{\ell\in T_v}\ \frac{2|\lambda_{\cC(\ell)}|}{  \cos^2\frac{\phi}2  }\ .
\eea
Finally, using $\sum_{\mathcal{T}_{v,\{{\mathcal{C}}(\ell)\}, i^*}} 1 = 7^{v-1}\frac{(2v-2)!}{(v-1)!}$ and $\frac{|\lambda_{\cC(\ell)}|}{  \cos^2\frac{\phi}2}\leq \rho$, we find absolute convergence of the series in the cardioid domain with $\rho = \frac1{56}$.

The perturbative expansion in $\lambda$ can be recovered by applying Taylor expansion with integral remainder individually on each tree, up to an uniform order. This corresponds to expanding beyond trees by adding additional loop-edges to the graph. This mixed expansion \cite{GurauT4} here does not have such simple properties as in \cite{Delepouve:2014bma}, because additional two-colored loops do not always lead to higher orders in $1/N$. The integral remainders however, still give a  bound on the overall Taylor remainder of $G_2$ which is good enough to prove Borel summability of the Feynman graph expansion uniformly in $N$. 

This also extends to higher-order cumulants, introducing trees with more than one cilium and Weingarten functions \cite{GurauT4,Delepouve:2014bma}. As in 
\cite{Delepouve:2014bma}, recovering proper scalings for the cumulants is done by Taylor expanding further than trees before using iterated Cauchy-Schwarz bounds. In this case however, two-colored loop edges do not lead to automatic scaling improvement, thus the expansion must be done with respects to single colored couplings only. This requires introducing several independent coupling constants. 

\section{Conclusion}

Pushing further the approach of \cite{b2} and based on the observation of \cite{universality} that non-melonic observables can be enhanced, we have studied two (related) families of models with \emph{maximally-rescaled} interactions.

First we introduced the most generic quartic model at rank four, which features four melonic interactions and three necklaces which are matrix-like observables. In addition to the existence of the $1/N$ expansion, we performed a combinatorial analysis of the leading order (LO) graphs in the large $N$ limit. In the case where only a single necklace has a non-vanishing coupling constant (called the restricted quartic case), Theorem \ref{thm:RestrictedLO} fully characterizes the LO graphs. They have the structure of trees whose vertices are disks, one tree being inserted on any face of the other one and so on.

The restricted case was then generalized to a generic class of interactions which we called trees of necklaces (the latter include all necklaces of a fixed color type as well as all melonic observables). We obtained a factorization theorem (Theorem \ref{thm:Factorization}) which reduces the large $N$ evaluation of the expectations of trees of necklaces. This, combined with Schwinger-Dyson equations, leads to the exact enumeration of the LO graphs contributing to the free energy or to such expectations. A phase transition (with a family of positive entropy exponents) is found between the branched polymer phase and 2D quantum gravity, which can be thought of as the proliferation of baby universes.

Beyond the specific outcome of the analysis (the Schwinger-Dyson equations reduce to the same equations as in matrix models with multi-trace invariants) we think the method we used is quite powerful and worth summarizing here.
\begin{itemize}
\item It starts with a quartic model. This is because for quartic interactions the intermediate field method is straightforward and powerful\footnote{See however \cite{Rivasseau:2010ke}
for the more complicated intermediate field representation of higher order interactions.}, simplifying greatly the combinatorial analysis. In particular, it was first used in \cite{Dartois:2013sra} to study a quartic tensor model beyond its LO and first sub-leading correction (resulting in the first double-scaling limit for random tensors). More recently, the intermediate field method for quartic models has been used to construct such models non-perturbatively.
\item Then a generic class of interactions is introduced: they are the boundary graphs of the quartic model studied before (in our case, of the restricted quartic model). This allows to consider the Feynman graphs of those new models as a sub-family of those of the quartic model. The combinatorial analysis performed in the quartic case thus applies.
\item That combinatorial analysis is then a fruitful input which simplifies the Schwinger-Dyson equations and can turn them into a solvable set of equations.
\end{itemize}
This scheme was in fact first applied in \cite{DSSD}. It enabled extending the double-scaling limit of the quartic melonic model of \cite{Dartois:2013sra} to arbitrary melonic models (invariant under permutations of the colors though). This is however the first time it is applied to non-melonic interactions. We emphasize that the Gaussian expectations of a sub-family of trees of necklaces were calculated recently in \cite{GaussianExpectations} (with a completely different method). Clearly, the present method goes far beyond the previous achievements.

The present results can be continued in several directions. First, it would be interesting to know the continuous spaces towards which the ``trees of disks'' converge in the GHP sense. It is tempting to think about them as generalization of the ``looptrees'' \cite{looptrees}, but further investigation is needed. The double-scaling limit of that model could also be studied to understand how the double-scaling limit of ordinary tensor models can cross over to the one of ordinary matrix models, and how it relates to the double-scaling limit of multi-trace matrix models \cite{KlebanovHashimoto}.

In this article, we left unsolved the full quartic model. Indeed, the presence of several matrix-like observables (necklaces of color types 12, 13 and 14) makes the Schwinger-Dyson equations more complicated. The fact they have different color types reveals that this model makes a better use of the tensorial structure. We expect it to depart from the multi-trace matrix models.

Another way to further take advantage of the richness of tensorial invariants is to introduce more maximally rescaled models.  Remark that already at rank 3 and order 6, in addition to the six melonic interactions, there is a single non-melonic tensor invariant, whose bubble is the bipartite complete graph $K_{3,3}$, which is non-planar and by Kuratowski's theorem, is a kind of canonical source for non-planarity in bipartite graphs (see Figure \ref{order6}). When maximally rescaled, it could generate an interesting leading order. This is left to future study. 

Finally it should be interesting to define similarly rescaled models of tensor group field theories, with or without Boulatov type projectors, to generalize the growing list of renormalizable models studied so far \cite{BGR, COR1, Sam1, COR2, BGO, BG1, BG2, BG3, BL}. In particular, they should allow to explore in depth the frontier between asymptotically free tensor field theories and asymptotically safe non-commutative field theories \cite{GW,Grosse:2004by,Disertori:2006uy,Disertori:2006nq,Grosse:2012uv}. 
 
\medskip

\noindent{\bf Acknowledgments}
We thank Razvan Gurau and Luca Lionni  for useful discussions.


\end{document}